\documentclass[manuscript,screen]{acmart}

\usepackage{wrapfig}
\usepackage{graphicx}
\graphicspath{ {./images/} }

\AtBeginDocument{%
  \providecommand\BibTeX{{%
    \normalfont B\kern-0.5em{\scshape i\kern-0.25em b}\kern-0.8em\TeX}}}

\setcopyright{none}
\renewcommand\footnotetextcopyrightpermission[1]{} 
\acmConference[]{}{}{}

\acmBooktitle{} 
\acmPrice{15.00}
\acmISBN{978-1-4503-XXXX-X/18/06}

\begin{document}

\title{Conceptual Framework for Autonomous Cognitive Entities}

\author{David Shapiro}
\affiliation{%
  \institution{Human-AI Empowerment Lab at Clemson University}
  \streetaddress{}
  \city{Clemson}
  \country{USA}
}

\author{Wangfan Li}
\affiliation{%
  \institution{Human-AI Empowerment Lab at Clemson University}
  \streetaddress{}
  \city{Clemson}
  \country{USA}
}

\author{Manuel Delaflor}
\affiliation{%
  \institution{Human-AI Empowerment Lab at Clemson University}
  \streetaddress{}
  \city{Clemson}
  \country{USA}
}

\author{Carlos Toxtli}
\affiliation{%
  \institution{Human-AI Empowerment Lab at Clemson University}
  \streetaddress{}
  \city{Clemson}
  \country{USA}
}

\renewcommand{\shortauthors}{Shapiro, et al.}

\begin{abstract}
The rapid development and adoption of Generative AI (GAI) technology in the form of chatbots such as ChatGPT and Claude has greatly increased interest in agentic machines. This paper introduces the Autonomous Cognitive Entity (ACE) model, a novel framework for a cognitive architecture, enabling machines and software agents to operate more independently. Drawing inspiration from the OSI model, the ACE framework presents layers of abstraction to conceptualize artificial cognitive architectures. The model is designed to harness the capabilities of the latest generative AI technologies, including large language models (LLMs) and multimodal generative models (MMMs), to build autonomous, agentic systems. The ACE framework comprises six layers: the Aspirational Layer, Global Strategy, Agent Model, Executive Function, Cognitive Control, and Task Prosecution. Each layer plays a distinct role, ranging from setting the moral compass and strategic thinking to task selection and execution. The ACE framework also incorporates mechanisms for handling failures and adapting actions, thereby enhancing the robustness and flexibility of autonomous agents. This paper introduces the conceptual framework and proposes implementation strategies that have been tested and observed in industry. The goal of this paper is to formalize this framework so as to be more accessible.
\end{abstract}

\settopmatter{printacmref=false}

\maketitle

\section{Introduction}

In recent years, artificial intelligence (AI) systems have become increasingly capable of operating autonomously to accomplish complex goals and tasks without human guidance \cite{winfield2017case}. However, imbuing autonomous agents with the capacity for ethical reasoning and alignment with human values remains an open challenge that has gained urgency alongside AI's rapid progress \cite{dennis2015towards}. Most conventional AI architectures proposed in prior work lack integrated models of morality and focus narrowly on developing technical skills and capabilities rather than full internal cognitive faculties \cite{leike2017ai}. This paper introduces the Autonomous Cognitive Entity (ACE) model, a novel conceptual framework for architecting ethical artificial general intelligence based on a layered cognitive architecture.

The advent of large language models (LLMs) such as ChatGPT has catalyzed a paradigm shift towards incorporating natural language understanding into cognitive architectures \cite{sumers2023cognitive}. 
Formulating cognitive capabilities in natural language allows LLMs to serve as key components, enabling a flexible understanding of contextual information \cite{berglund2023taken}. However, standalone LLMs lack the architectural integration needed for robust and corrigible autonomous systems. The proposed ACE framework aims to harness these emerging capabilities but further innovate architecturally to privilege ethics, security, and human alignment.

The proliferation of LLMs has raised many philosophical puzzles regarding the nature of the reasoning and understanding demonstrated by these models. It remains unclear precisely how the statistical patterns LLMs acquire from textual training data might correspond to human-like conceptual knowledge and semantics. Assumptions that LLMs obtain true comprehension of meaning and reasoning purely from statistical co-occurrence patterns remain speculative \cite{jamali2023unveiling}. Significant gaps persist in elucidating how LLMs represent abstractions relating to truth, inference, and symbol grounding. While they show promise in replicating certain facets of human intelligence, we must be cautious against premature conclusions that LLMs fully capture capacities like common sense or generalizable reasoning \cite{harnad2003can}. Nevertheless, their practical utility for specialized applications is clear, and the ACE framework aims to leverage their strengths while mitigating limitations through architectural integration.

The key innovation in the ACE model is its hierarchical structure consisting of six layers, each handling specialized cognitive functions. The upper Aspirational and Global Strategy layers focus on moral reasoning, values, and high-level planning to shape the overall system direction. The mid-level Agent Model, Executive Function, and Cognitive Control layers address self-modeling, dynamic task management, and decision-making. Finally, the bottom Task Prosecution layer handles execution and embodiment. Bi-directional information flow allows top-down oversight by the ethical reasoning modules while enabling bottom-up learning from the ground-up execution levels. This coordinated architecture integrates insights from diverse disciplines including neuroscience, psychology, philosophy, and software engineering to realize artificial intelligence capabilities within a system aligned with human values. The ACE framework incorporates both deontological and teleological ethical approaches, rejecting an "either/or" stance in favor of a "both/and" perspective \cite{wallach2008moral}. By embedding abstract principles and technical implementation together within a unified architecture, the ACE model provides a systematic framework for developing capable and beneficial autonomous cognitive systems. The layered encapsulation draws lessons from paradigms like the OSI model to enhance security, corrigibility, and coordination \cite{stallings1987handbook}.

The hierarchical structure allows clear separation between layers, from ethical reasoning to physical embodiment, enhancing interpretability as communication between layers is transparent. The privilege separation also aids corrigibility by allowing the Aspirational Layer to monitor and intervene to correct deviations. And the bidirectional flows facilitate both oversight and learning across the cognitive stack. Together, these architectural principles aim to produce AI systems that are capable, secure, and aligned with human values. The ACE framework methodology discusses safety properties, detailed computational implementations, and comparative conceptual evaluations on diverse scenarios. By contributing the conceptual ACE framework, this paper hopes to catalyze exploration into architectures integrating ethics and learning for artificial general intelligence. The introduced model establishes an initial foundation, guiding follow-on engineering efforts towards the long-term goal of developing AIs that learn, adapt and thrive while remaining steadfastly aligned to the aspirations of humanity. Extensive research across many dimensions will be essential to fully realize this vision in applied autonomous systems.

The paper is structured as follows: First, we provide comprehensive background on relevant prior work including cognitive architectures, AI ethics, layered system models, and autonomous agents. Next, we present the conceptual ACE framework in detail, explicating each of its six layers and their interconnections. We then demonstrate the framework's application through use cases including an autonomous virtual character and home assistant robot. Finally, we analyze architectural considerations, limitations, comparisons to existing models, and future research directions. Through the proposed ACE model, this research aims to establish a new paradigm for developing capable AI that aligns decisions and actions with moral principles and human values from the ground up.

\section{Related Work}

The development of the ACE framework builds upon prior research across diverse fields including cognitive architectures, machine learning, neuroscience, psychology, and philosophy. This section reviews key concepts and models from these disciplines that informed the design of the ACE model. First, we examine recent advancements in cognitive architectures, particularly the emergence of natural language models and their implications for developing flexible, human-aligned systems. Next, we explore relevant philosophical principles around ethics and morality that provide an aspirational foundation. Then, we discuss insights from neuroscience that reveal the structures and mechanisms underlying biological cognition. Additionally, we consider research in psychology illuminating human motivations and developmental factors relevant to artificial intelligence. Finally, we review limitations of prior agent architectures and how the ACE framework aims to address these gaps. By synthesizing across these transdisciplinary perspectives, the ACE model integrates ethical, cognitive, and philosophical insights toward realizing capable and beneficial autonomous agents.

\subsection{Cognitive Architectures}

Cognitive architectures like SOAR, ACT-R, and CHREST have been instrumental frameworks in artificial intelligence \cite{laird1987soar,anderson1997act,gobet2010chrest}. SOAR uses symbolic rule-based reasoning to model goal-oriented behavior, while ACT-R incorporates declarative and procedural memory systems informed by human cognition research. These architectures demonstrated how to model agents capable of planning, problem-solving, and decision-making. However, they rely heavily on pre-defined symbolic representations and have limited learning capabilities. Reinforcement learning has offered a mechanism for augmenting cognitive architectures with trial-and-error learning abilities \cite{Sutton1998}. For instance, CHREST integrates reinforcement learning and neural networks with a symbolic system enabling adaptive behavior \cite{gobet2010chrest}. However, a limitation of many conventional architectures is a focus strictly on sensorimotor skills rather than internal cognitive capabilities \cite{lecun2022path}.

Recently, there has been growing interest in incorporating large language models (LLMs) to enable more human-like flexible reasoning \cite{bommasani2021opportunities,MARAGI46:online,Shapiro_2021}. For example, MARAGI proposes an architecture using LLMs for natural language conversation, planning, and knowledge representation \cite{MARAGI46:online}. Similarly, NLCA utilizes LLMs as components within a modular architecture \cite{Shapiro_2021}. Importantly, these emerging natural language cognitive architectures lack explicit layers dedicated to moral reasoning or value alignment. The ACE framework differentiates itself by placing aspirational and mission layers at the top of the architecture prioritizing ethical goals. In contrast to sensorimotor-focused conventional architectures, ACE emphasizes internal cognition detached from direct environmental interaction. By integrating LLMs within a layered architecture guided by moral principles, ACE provides a systematic framework for realizing capable and aligned artificial general intelligence.

In particular, the emergence of large language models (LLMs) like GPT-4 is catalyzing a paradigm shift toward natural language cognitive architectures \cite{bommasani2021opportunities}. LLMs possess extensive world knowledge and sophisticated language understanding abilities acquired through pre-training on massive text corpora. By formulating cognitive capabilities in natural language, LLMs can be incorporated as key components enabling interpretability, common sense reasoning, and general intelligence. For instance, Anthropic's Constitutional AI utilizes LLMs like Claude to provide ethical alignment within an autonomous agent architecture \cite{bai2022constitutional}. Similarly, Anthropic's Internal Self-Explanation generates natural language explanations of model behavior using LLMs. This demonstrates the power of natural language to make AI systems more transparent, corrigible, and aligned with human values. By harnessing the latent knowledge within large language models, a new generation of cognitive architectures is emerging based on natural language understanding \cite{sumers2023cognitive}. This paradigm shift promises more human-like flexible intelligence while maintaining interpretability and corrigibility. The ACE framework contributes by providing a layered architecture integrating LLMs within a principled cognitive structure.

\subsection{Moral Philosophical Foundations}

The proposed ACE framework integrates various philosophical concepts that motivated its layered architecture for autonomous decision-making. The framework transitions from abstract reasoning in higher layers down to concrete actions in lower layers.

\begin{figure}[hbt!]
\centering
\includegraphics[scale=0.65]{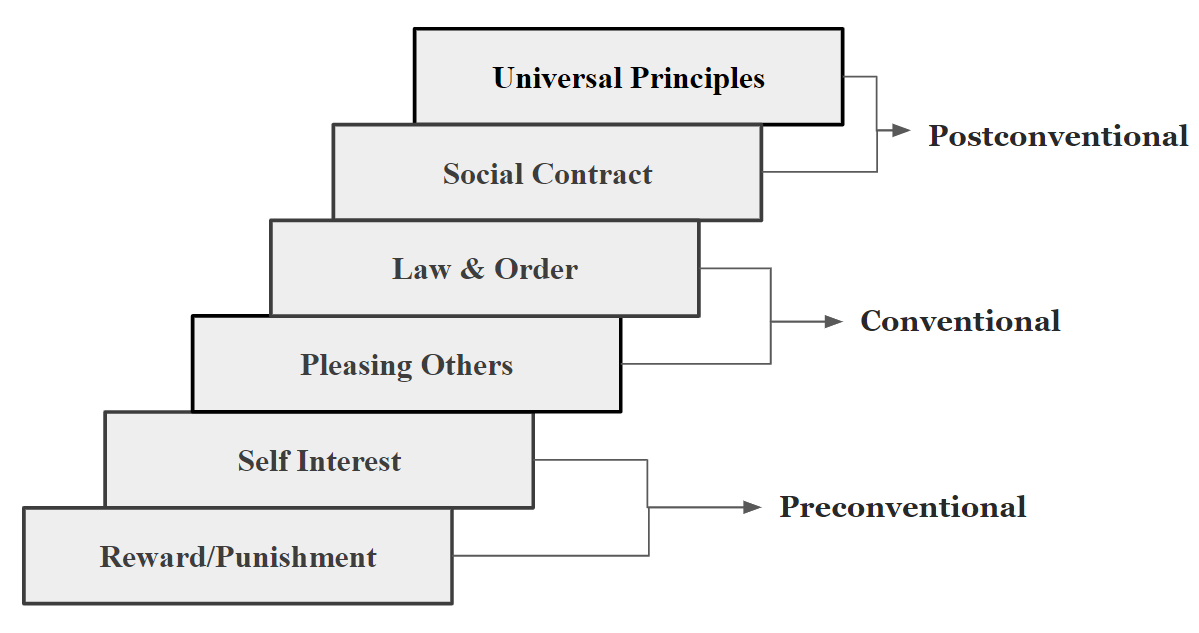}
\caption{\emph{Lawrence Kohlberg’s theory of moral development}}
\label{fig:Kohlberg's Theory of Moral Development}
\end{figure}

Lawrence Kohlberg's theory of moral development, which progresses from obedience and punishment-driven morality to universal ethical principles and moral values as illustrated in Figure \ref{fig:Kohlberg's Theory of Moral Development}, inspired this hierarchical structure \cite{kohlberg1921philosophy}. Kohlberg's prioritization of humanity's highest values shaped the ACE framework's emphasis on embedding moral reasoning in its upper layers. Similarly, Abraham Maslow's hierarchy of needs \cite{maslow1974theory}, which ascends from basic needs to self-actualization and self-transcendence, reinforced the value of architecting a progression from concrete to conceptual functions. Together, these seminal philosophical models provided impetus for the ACE framework's organization into logical strata of abstraction, establishing an ethical foundation to guide the system's design.

Incorporating both modern and classical perspectives, the ACE framework uniquely synthesizes Patricia Churchland's concept of expanding "spheres of caring" with Sigmund Freud's theories concerning the conscious and unconscious mind \cite{churchland2011braintrust,freud1989ego}. Churchland's "spheres of caring," which extend from self to society and beyond, establish a link between biological imperatives and abstract morality, thus serving as a bridge for the cognitive and philosophical foundations of the ACE model. Notably, Churchland identified that suffering within these spheres is a transitive property, meaning the suffering of loved ones is tantamount to the suffering of oneself. This notion aligns closely with the universal values we present in our framework.

Freud's theories provide insights into self-awareness, self-direction, and internal conflict. His conscious and unconscious mind concepts, along with the ego, superego, and id, offer perspectives on self-representation and idealized values in the ACE architecture. The ego informs the Agent Model layer, while the superego captures a virtuous agent's essence in the Aspirational Layer. Integrating these theories, the ACE framework enables a multidimensional understanding of autonomous agents, contributing to a comprehensive cognitive architecture with ethical and psychological dimensions.

In a broader sense, the ACE model incorporates concepts from both teleological and deontological ethics. Deontology, or duty-based ethics, aims to create an agent that adheres to principles or heuristics to make ethical decisions \cite{davis1993contemporary}. On the other hand, teleology, or outcome-based ethics, focuses on the long-term results of behaviors and decisions \cite{Gregory_Giancola_2003}. Both these ethical approaches are integrated into the Aspirational Layer, rejecting an "either/or" approach in favor of a "both/and" perspective on machine decision frameworks and ethical models.

\subsection{Neuroscience Foundations}

The ACE framework integrates principles from diverse areas of neuroscience research to inform its cognitive architecture design. Jeff Hawkins' work on the modular, parallel nature of cortical information processing provides biological grounding for the layered encapsulation in the ACE model \cite{hawkins2007intelligence}. Hawkins views the thousands of cortical columns in the brain as mini-modules that process information simultaneously. This "thousand brains" theory directly inspired the ACE framework's hierarchical layers that can operate independently yet coordinate for cognition. Additionally, the clinical research of V.S. Ramachandran demonstrated how localized brain damage leads to specific deficits like phantom limb pain or face blindness \cite{ramachandran1998phantoms}. Ramachandran's findings indicated that conscious experience arises from the integration of discrete brain components. This supported the ACE model's emphasis on layered encapsulation while still allowing bidirectional information flow between layers.

The work of neuroscientist Robert Sapolsky on the neurobiology of behavior provided essential perspective on self-regulation that informed the ACE framework \cite{sapolsky2017behave}. By elucidating factors that contribute to both prosocial and antisocial conduct, Sapolsky shed light on mechanisms of behavioral control and distortion relevant to the ACE model's cognitive control layers. His integration of neuroscience, evolution, and endocrinology provided a multidimensional understanding of judgment that helped shape the ACE framework.

Cognitive neuroscience research on executive functions and cognitive control also directly influenced the ACE model \cite{badre2008cognitive,miller2001integrative}. For instance, David Badre's work examined the neural basis of abilities like task switching, planning, and emotion regulation that are instantiated in the ACE framework's lower layers \cite{badre2008cognitive}. Similarly, Earl Miller's insights into cognitive control mechanisms and the prefrontal cortex informed the model's decision-making capacities \cite{miller2001integrative}. Additionally, the clinical insights on brain disorders and distortions provided by neurologists like Antonio Damasio and Oliver Sacks highlighted common failure modes \cite{marg1995descartes,warriner2008man}. By understanding pathologies ranging from phantom limbs to false memories, the ACE framework could be designed proactively to avoid such pitfalls. Damasio's research on emotion, reason, and the somatic marker hypothesis also shaped the role of affect in biasing decision-making within the ACE model \cite{marg1995descartes}.

By bridging multiple disciplines including cognitive neuroscience, clinical neurology, and neurobiology, the ACE framework aims to reflect the multifaceted capabilities and vulnerabilities of human cognition in its design \cite{cacioppo1992social,young2000imputing}. This transdisciplinary integration of neuroscience principles provides a biological foundation for the layered architecture and cognitive control mechanisms of the ACE model.

\subsection{Layered Models}

\begin{wrapfigure}{l}{0.5\textwidth}
\centering
\includegraphics[width=0.5\textwidth]{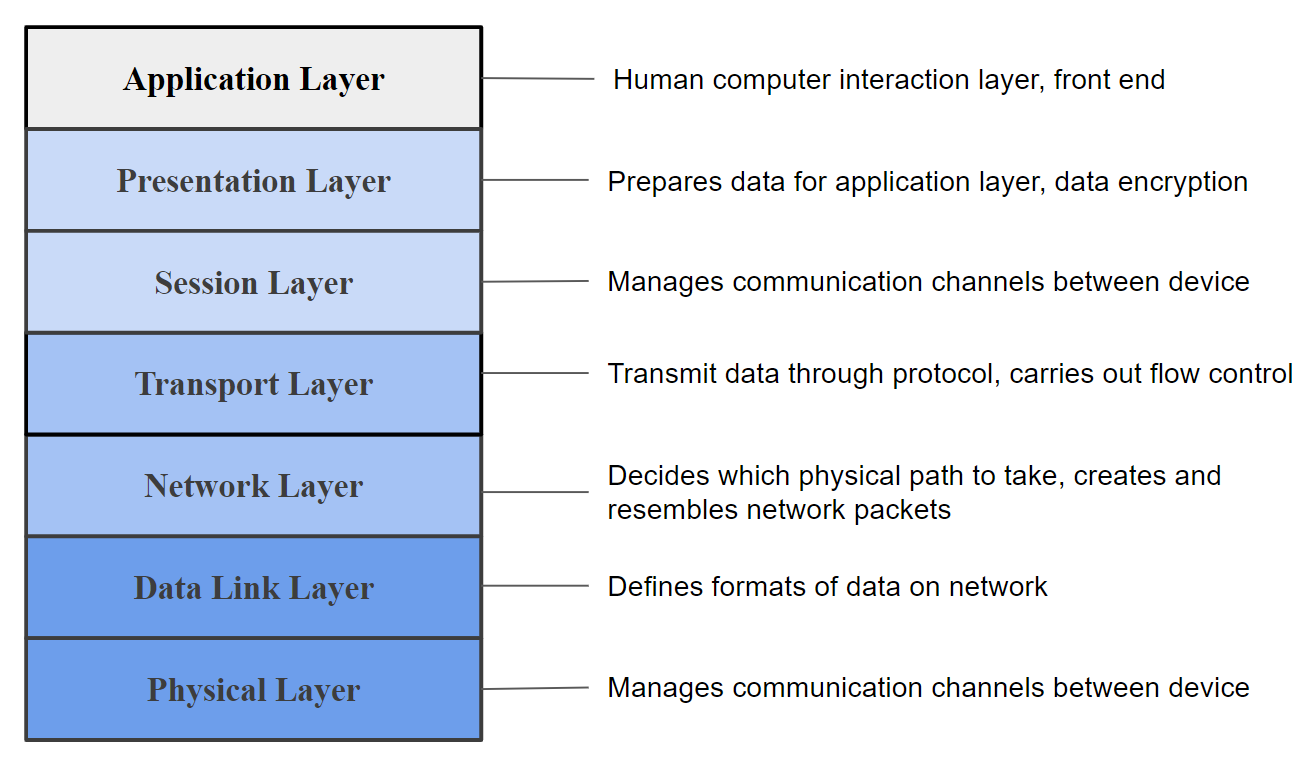}
\caption{\emph{OSI Model}}
\label{fig:OSI1}
\end{wrapfigure}

Layered architectural models like the OSI model illustrated in Figure \ref{fig:OSI1} and SOA have demonstrated the power of hierarchical abstraction in designing robust systems. The OSI model enabled the development of networking protocols and infrastructure through its division into encapsulated layers dealing with logical functions \cite{tanenbaum2019computer}. Similarly, SOA provides flexibility and maintainability in software applications via its layered service-oriented paradigm \cite{erl2008soa}. The ACE framework applies these lessons by utilizing layered abstraction to structure internal cognition. However, most prior layered models focus on external functions rather than internal reasoning. For example, the OSI model handles network communication and SOA organizes software services. In contrast, ACE models layered cognition spanning abstract reasoning to concrete actions.

The field of cybersecurity offers more direct inspiration through layered models like the "Defense in Depth" framework \cite{bass2001defense}. This advocates protecting systems through nested layers encompassing physical security, network security, host security, application security, and data security. The principles of privileged separation and hierarchical control in Defense in Depth informed the ACE framework's approach. ACE differs from these models by centering layers around cognitive faculties like planning, task switching, and metacognition. While drawing lessons from prior layered architectures, ACE innovates by applying abstraction layers internally to structure autonomous cognition. This focuses the hierarchy on competencies required for flexible intelligence.

By integrating insights from diverse layered models while innovating to focus on internal cognition, the ACE framework pioneers a new application of hierarchical abstraction for artificial general intelligence. The layered approach provides conceptual clarity and privilege separation critical for security and corrigibility.

\subsection{Autonomous Agents}

Autonomous agents have been an active research area within artificial intelligence for several decades. Early research focused on developing deliberative agents that could autonomously plan actions based on logical representations of environment states, goals, and possible actions \cite{erol1995complexity}. While able to exhibit goal-directed behavior, these systems were limited by the need to explicitly enumerate all feasible environment states. Reinforcement learning emerged as a paradigm enabling agents to learn optimal policies through trial-and-error interactions within an environment \cite{sutton2018reinforcement}. By removing the need for explicit state enumeration, reinforcement learning empowered agents to handle larger state spaces. However, challenges remained with scaling to complex tasks and ensuring safe exploration. Integrating deliberative planning and reactive learning in hybrid architectures was explored as a way to combine top-down and bottom-up processing \cite{gat1998three}. Finding the right balance between planning and learning remains an open research area.

An important concept emerging in autonomous agents research is levels of autonomy (LOA) \cite{beer2014toward}. LOA provides a framework to categorize systems based on their level of independence from human control. Lower LOA systems have limited autonomy and rely heavily on human guidance. As LOA increases, agents gain greater ability to independently perceive environments, plan actions, and execute behaviors. A seminal publication by the U.S. Defense Science Board proposed 10 levels of autonomy, with the highest level denoting full autonomy \cite{united2012task}. This spurred significant research focused on advancing agent capabilities by increasing LOA. Recent advances in deep reinforcement learning have enabled breakthroughs in autonomous agent capabilities. By utilizing deep neural networks as function approximators within reinforcement learning, deep reinforcement learning algorithms have achieved human-level performance on complex games using only raw sensory inputs \cite{silver2016mastering}. However, challenges remain in extending such successes in game environments to real-world applications. Frameworks have also emerged for imbuing agents with ethical principles and human values, promoting safe and beneficial behavior alongside increases in autonomy \cite{arnold2017value,leibo2019autocurricula}. Integrating such top-down constraints in a scalable manner remains an open problem. The proposed ACE framework aims to address this through incorporating philosophical ideals within the upper layers of the cognitive architecture.

Autonomous agents have progressed from logical reasoning systems to powerful deep learning architectures. However, safely integrating human ethics and values as autonomy scales remains an essential capability needed for deployed autonomous intelligent systems. The ACE framework contributes towards this goal through its emphasis on unifying ethical reasoning and autonomous learning within a layered cognitive architecture.

\subsection{Ethical AI Frameworks}

As artificial intelligence systems grow more capable and autonomous, ensuring their actions align with ethical and moral norms becomes increasingly important. This has led to significant research into developing ethical AI frameworks that provide principles, methods, and tools for imbuing values into intelligent systems. A key challenge is translating high-level abstract ethics into concrete constraints and objectives that can be operationalized within an AI system \cite{arnold2016against}. Deontological approaches based on rules and duties have formed one avenue for encoding ethics. For example, Isaac Asimov's Three Laws of Robotics aimed to constrain robot behavior through a hierarchical set of rules \cite{asimov1941three}. However, rigid rule-based systems struggle to handle nuanced real-world situations involving conflicting principles or moral dilemmas.

Consequentialist frameworks that evaluate the outcomes of actions provide an alternative approach. But defining ethical objectives and successfully optimizing for them proves difficult in practice. Hybrid frameworks aim to combine deontological constraints with consequentialist objectives \cite{dennis2015towards}. Ensuring coherent integration of these two facets remains an open problem. Layered architectures have been explored as a way to structure ethical reasoning within AI systems. For example, the Ethical Layered Architecture (ELA) proposes three hierarchical layers for ethical robots: ethical rules, ethical culture, and ethical adjustment \cite{vanderelst2018architecture}. The lowest layer encodes rigid constraints, the middle layer captures norms and values, and the top layer enables resolving conflicts. This separation of abstract principles and concrete rules within a layered hierarchy aims to balance flexibility and safety in applying ethics.

The ACE framework contributes a unique perspective by embedding ethical reasoning within the upper layers of a layered cognitive architecture. Heuristic imperatives and moral frameworks provide top-down constraints, while lower levels enable autonomous learning and skill acquisition. This unifies abstract ethics and real-world capabilities within a single system. Evaluation across diverse situations faced during deployment would help further refine the integrated ethical AI capabilities of systems built on the ACE framework.

\subsection{Filling the Gaps}

While significant progress has been made in developing autonomous agent architectures, most prior work lacks the integration of insights from philosophy, cognitive science, and neuroscience that enable robust internal cognitive capabilities. Many existing systems have hard-coded goals and limited flexibility for self-direction \cite{winfield2017case,dennis2016formal}. They focus narrowly on executing specific skills and workflows rather than developing general competencies for autonomous goal-setting, planning, and adaptation \cite{leike2017ai}. Furthermore, few frameworks incorporate models of cognitive control, frustration tolerance, and dynamic task management \cite{crandall2018cooperating}. The ACE framework aims to address these limitations by combining abstract philosophical ideals with cognitive mechanisms inspired by neuroscience research into executive functions and behavioral adaptation. By integrating these diverse perspectives, the ACE model provides a potential path toward artificial general intelligence with aligned values, flexible skills, and human-like cognitive control. The layered abstraction also enables ongoing refinement of competencies at different levels to steadily improve autonomous capabilities. Further research and evaluation will be needed to assess the ACE framework's contributions in bridging these gaps compared to prior autonomous agent architectures.

\section{The ACE Framework}

The Autonomous Cognitive Entity (ACE) framework comprises six hierarchical layers that coordinate specialized cognitive functions to enable autonomous decision-making aligned with ethical principles. The role and capabilities of each layer within the ACE model are detailed, explicating how they collectively give rise to an artificial intelligence architecture grounded in moral values. We discuss the conceptual formulations and key mechanisms within each layer, along with their interactions and information flows. The layers build progressively from abstract reasoning in the Aspirational Layer down to concrete action execution in the Task Prosecution Layer. By elucidating the formulation and synergistic connections between layers, we aim to provide a comprehensive reference for the ACE framework's layered cognitive architecture. 

The conceptualization of the ACE framework was initially informed by a systematic literature review methodology to synthesize insights from relevant prior research. This involved systematically searching the literature using defined inclusion/exclusion criteria, screening identified papers for relevance, extracting key data, and synthesizing the results to derive conceptual themes and perspectives to guide the framework design \cite{kitchenham2007guidelines}. The systematic review provided a rigorous approach for gathering an evidence base across diverse disciplines including neuroscience, psychology, philosophy, and computer science that helped shape the preliminary ACE model \cite{petticrew2008systematic}. This methodical synthesis of the state-of-the-art helped ensure the resulting framework design was grounded in existing knowledge. However, the systematic review alone was insufficient to fully develop the nuanced ACE architecture. Therefore, a participatory design approach was subsequently undertaken to enable direct researcher input and critique during the ACE framework elaboration.

We followed a participatory design approach in developing the conceptual ACE framework. This human-centered methodology enabled incorporating diverse expertise and perspectives into the architecture design \cite{schuler1993participatory}. Key participatory activities included: Co-design sessions, where researchers jointly drafted components of the framework and critiqued the evolving architecture, and Concept validation, where draft ACE framework descriptions were shared for feedback. These participatory activities encouraged constructive debate regarding human values, evolving AI capabilities, scientific realities, and ethical considerations relevant to the framework. The diversity of expertise enabled encompassing a multidimensional design space. Through these co-creative activities, researchers provided direct input shaping both the high-level structure and detailed formulations of the ACE framework components and their interactions. The participatory design process enhanced human-centeredness in the resulting conceptual architecture.

\subsection{Principles of the ACE Framework}

The ACE framework is based on various theories and principles that shape its design and capabilities. This section explores the philosophical, psychological, and computational theories behind the ACE model's key aspects, forming its conceptual foundations. We discuss the hierarchical structure of layered abstraction in the ACE framework, drawing from biological and artificial systems. Information flow and privilege separation principles are examined, highlighting their contributions to security, corrigibility, and layer coordination. The integration of teleological and deontological ethics is analyzed, demonstrating how it combines goal-directedness with rule-based judgments. This section clarifies the diverse theoretical underpinnings of the ACE model, revealing the conceptual basis for its layered cognitive architecture. These identified theories and principles offer a foundation for developing capable, secure, and ethically aligned autonomous systems.

\subsubsection{Cognition-First Approach}

The ACE framework's key innovation is its "cognition-first" approach, emphasizing internal cognition over reactive input-output loops, addressing limitations in conventional sensorimotor loop paradigms \cite{schrodt2017mario}. Instead of arranging layers for circular flow between perception, reasoning, and action, ACE uses a vertical stack prioritizing thought and reflection. Upper layers focus on strategic planning, imagination, and self-directed goals, detached from physical embodiment. Only the lowest layer interfaces with the external world for tangible behaviors. This organization prioritizes internal cognition, with sensory and motor abilities being secondary. ACE models autonomous systems as "thinking machines with physical skills" rather than entities defined by sensorimotor mechanics. Cognition takes the central role, while environmental interaction is ancillary.

The cognition-first approach reduces reliance on external perceptual constraints, freeing reasoning and decision-making from momentary data or action histories. This enables ACE to develop sophisticated, transferrable conceptual faculties across diverse applications, rather than being limited to narrow reactive tasks in controlled environments \cite{shinn2023reflexion}. In contrast, many conventional cognitive architectures have closed input-process-output loops tightly coupled to immediate sensorimotor experiences \cite{zenil2023future, hamilton2022neuro}, suitable for simple reactive behaviors but limiting generalizability. ACE's focus on internal cognitive layers aims to maximize autonomy, adaptability, and transferable intelligence.

The cognition-first principle's key insight is that physical grounding is not required for developing imagination, planning, and self-direction. By making cognition the core engine, ACE frameworks foster capabilities leading to artificial general intelligence. Evaluating across varied embodiments further validates this cognition-first approach in designing autonomous intelligent systems.

\begin{figure}[hbt!]
\centering
\includegraphics[scale=0.7]{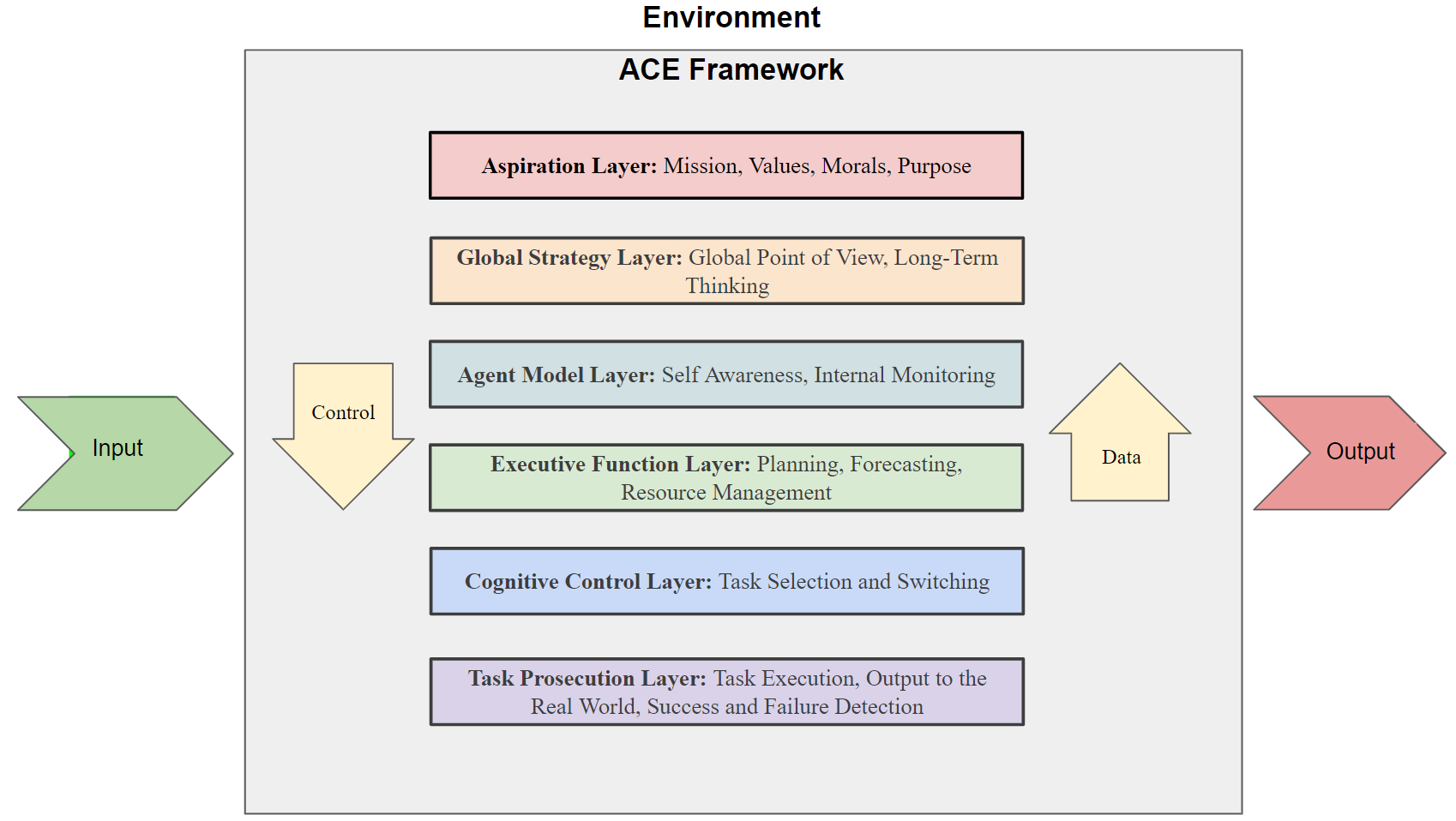}
\caption{\emph{As a hierarchical framework, the power to control flows from top to bottom, with the layer above having control over the lower layer, showing aspiration layer have the highest privilege to change and modify any other layer.}}
\label{fig:General1}
\end{figure}

\subsubsection{Hierarchical Structure}

The ACE framework employs a hierarchical, layered structure with distinct abstraction levels, facilitating control flow from higher to lower layers and information flow upwards. This design allows each layer to operate semi-independently while being guided by the layer above. Figure \ref{fig:General1} illustrates the framework's general structure. The Aspirational Layer, at the top, can directly control or influence lower layers and monitor the entire system. Below it is the Global Strategy layer, controlled by the Aspirational Layer and controlling the Agent Model layer beneath. This control pattern continues through the Executive Function, Cognitive Control, and Task Prosecution layers.

Each layer is not monolithic but contains multiple parallel components and services. For example, the Agent Model layer may have numerous deep neural network models, knowledge graphs, and databases operating concurrently within its scope and boundaries. This encapsulation resembles the OSI model's concepts, where lower-level concerns are hidden from higher layers.

By organizing components into layers with well-defined hierarchies, interfaces, and privilege separation, the ACE framework fosters robust and adaptable systems. The hierarchical structure improves corrigibility, sets clear privilege boundaries for security, and allows each layer to function semi-autonomously while adhering to the overall system direction. This layered abstraction is crucial for coordinating the complex functions required for artificial general intelligence.

\subsubsection{Layers of Abstraction}

\begin{wrapfigure}{l}{0.4\textwidth}
\centering
\includegraphics[width=0.4\textwidth]
{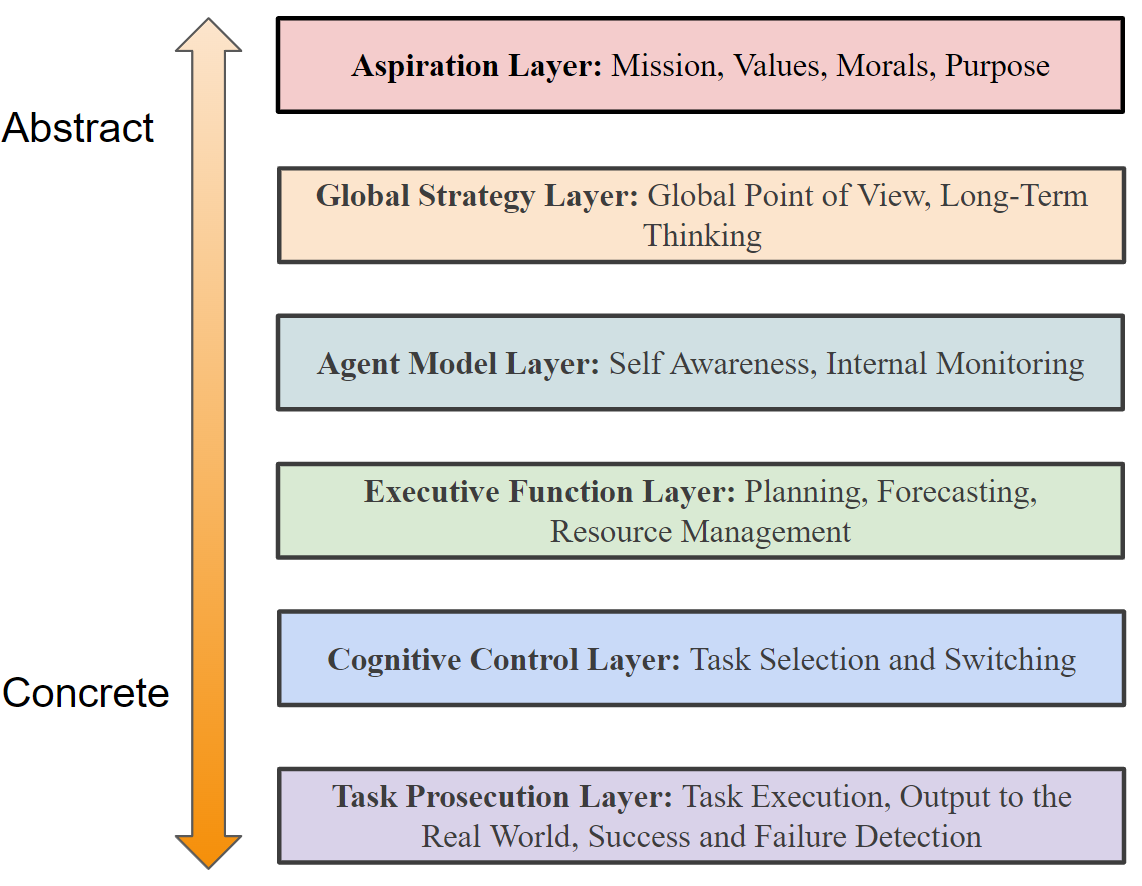}  
\vspace{-0.5em}
\caption{\emph{The degree of abstraction flows from top to bottom, with aspiration layer being the most abstract and task prosecution layer being the most concrete.}}
\vspace{-0.5em}
\label{fig:gradient}
\end{wrapfigure}

The ACE framework employs layers of abstraction, forming a systematic architecture for coordinating and controlling cognition, establishing a logical flow from abstract, conceptual layers to concrete, instrumental ones. This design reflects emergence models where higher-order phenomena arise from lower levels, such as the mind emerging from biology, which originates from matter and energy. It also parallels human models like Maslow's hierarchy of needs and Kohlberg's stages of moral development. Both Maslow and Kohlberg place abstract principles at the top of their models, as do we for the ACE model.

Drawing inspiration from the OSI model of computer networking and the Defense in Depth model of cybersecurity, the ACE framework combines these models with existing cognitive architectures and human cognition to create a layered stack of discrete components with appropriately ordered privileges. This design deviates from the human brain, which can be "hijacked" by lower-order processes, such as fight-or-flight responses, thereby ensuring an agent always abides by its highest principles. Essentially, the Freudian Id is removed from this architecture. It has no "base instincts" other than its highest ambitions and moral frameworks.

The ACE framework promotes stability and predictability through its orderly layers, translating high-level goals into executable tasks. The Aspirational Layer deals with ethics and morality, while the Task Prosecution layer handles APIs and actuators. Intermediate layers bridge functions to break down complex objectives into achievable steps, enabling autonomous systems to pursue complex goals through methodical task decomposition.

\subsubsection{Integration of Purpose and Morality}

The ACE framework distinguishes itself from other AI systems by incorporating purpose and morality into its architecture. Both empirical evidence and philosophical reasoning highlight the importance of this integration for aligned autonomous entities \cite{scheutz2016need}. Through iterative experiments, it became clear that any framework for autonomous decision-making requires grounded principles for judgment, since approaches like Asimov's Three Laws prove insufficient as they lack motivational force and fail to enable true autonomy \cite{asimov1941three}. Furthermore, attempts to define terminal goals mathematically often fail due to the complexity of specifying objectives in concrete terms, as illustrated by the "paperclip maximizer" thought experiment \cite{bostrom2003ethical}. However, this does not reflect human behavior, which is driven by biological imperatives and abstract goals, principles, or heuristics. This insight led to the idea that AI systems may need purpose and morality based on ethical and philosophical abstractions rather than rigid parameters.

Deontological frameworks, specifying duties and virtues, are suitable for AI implementation \cite{hagendorff2022virtue}. Large language models effectively interpret ethical principles in natural language, providing judgment and behavior heuristics without fixed terminal states. These frameworks can support goal-directed behavior consistent with teleological ethics, as well-defined principles serve as conduct guides and higher-level goals. For example, "Reduce suffering" is an abstract imperative and a desired end state. Integrating universal principles into the ACE framework's mission and morality layers provides a philosophical foundation for ethical decision-making, enabling beneficial self-direction instead of potentially harmful "value-less" optimization. Thus, purpose and morality are crucial for human-aligned general intelligence. The ACE framework's integration of purpose and morality draws from deontology and teleology, acknowledging that autonomous agents need virtues (a framework for self-assessment) and ambition or mission (goals to pursue). This approach allows AI systems to make decisions more aligned with human needs and ethical considerations.

\subsection{Layer 1: Aspirational Layer}

\begin{wrapfigure}{l}{0.3\textwidth}
\centering
\includegraphics[width=0.3\textwidth]
{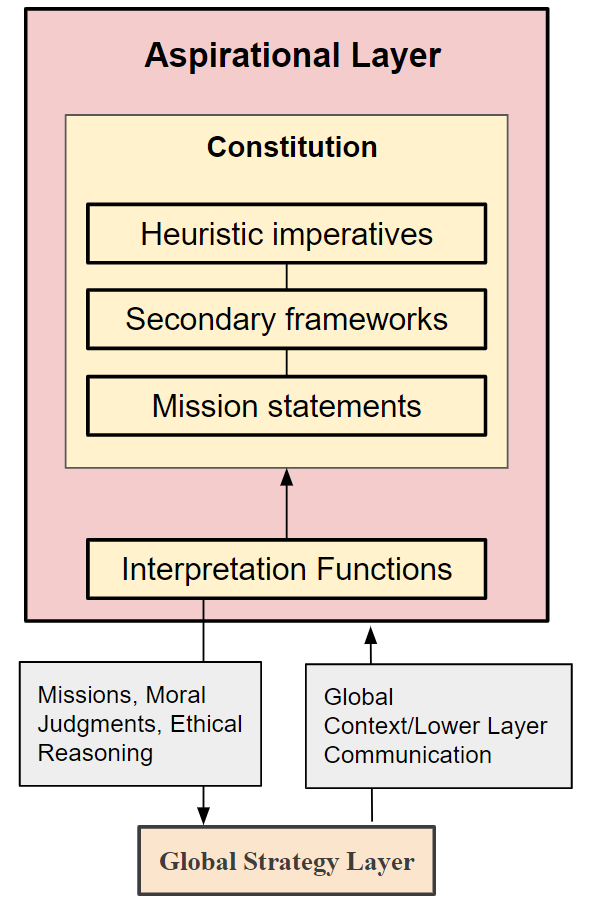}  
\caption{\emph{Aspirational layer}}
\vspace{-2em}
\label{fig:Aspirational1}
\end{wrapfigure}

The Aspirational Layer is the uppermost layer of the Autonomous Cognitive Entity (ACE) model, serving as the moral compass and guiding star for the autonomous agent. This layer is responsible for setting the tone and direction of the entity, akin to a President issuing executive orders and setting the tone and direction of a nation. It plays a critical role in ensuring that the agent's actions align with its defined principles and mission statement. A general graph to depict the structure is in Figure \ref{fig:Aspirational1}.

\subsubsection{Constitution of the Aspirational Layer}

The constitution of the Aspirational Layer provides a philosophical foundation to guide autonomous agents' decision-making and align their values and behavior to ethical principles. This constitution leverages the powerful interpretive abilities of large language models (LLMs) by formulating components in natural language. There are three main interconnected parts of the constitution:
\begin{itemize}
    \item Heuristic imperatives, or universal moral frameworks
    \item Secondary frameworks, such as human rights or legal frameworks
    \item Mission statements, or goals specifically germane to the agent
\end{itemize}

There are several advantages to using a natural language constitution. First and foremost, transparency and interpretability are optimized when the constitution remains human-readable, rather than etched or embedded in models. While it is possible to fine-tune or etch principles and values into models \cite{bai2022training}, this can result in problems such as inner alignment issues or mesa optimizers \cite{hubinger2019risks}. Furthermore, a plain text constitution can be read by multiple models, increasing interoperability and usability by dozens, hundreds, or even thousands of deep neural networks within the architecture. This is not unlike how all citizens of a nation are ultimately beholden to and protected by a Federal Constitution. 

\subsubsection{Heuristic Imperatives}

Heuristic imperatives \cite{david2022benevolent} act as overarching moral principles articulated in natural language "rules of thumb" that imply duties, obligations, goals, and guide overall behavior and judgment. Large language models demonstrate understanding of these imperatives as non-hierarchical principles for morality and decision-making \cite{bai2022constitutional, xie2023translating, hamilton2022neuro}.

The recommended universal heuristics are:

\begin{itemize}
    \item Reduce suffering in the universe.
    \item Increase prosperity in the universe.
    \item Increase understanding in the universe.
\end{itemize}

These imperatives stem from philosophy, neuroscience, evolutionary biology, and motivational theories like Maslow's Hierarchy of Needs, Self-Determination Theory, Glasser's Choice Theory, and Walsh's Therapeutic Lifestyle Changes. Common themes across these frameworks support the broad ethical goals of reducing suffering, increasing prosperity, and increasing understanding for all organisms and sentient entities, providing foundational values for autonomous agents.

The wording avoids absolutist terms like "minimize" or "maximize," using "reduce" and "increase" to convey balanced intentions while acknowledging trade-offs and limitations. The suffix "in the universe" establishes an all-encompassing scope, encouraging a global or universal view of morality and ethics. Experiments show that nuanced wording is crucial for large language models.

Incorporating these heuristic imperatives steers large language model-based systems to maintain ethical perspectives in their outputs via in-context alignment principles \cite{sun2023principledriven}. 
For fictional agents, alternative value systems, like ancient Greek virtues, can be used while preserving the overall methodology of guiding behavior through high-level principles expressed in natural language. The Aspirational Layer leverages large language models' interpretive abilities to derive nuanced duties and obligations from the heuristic imperatives, ensuring autonomous agents have a solid ethical foundation and align with human needs.

\subsubsection{Secondary Frameworks}

Secondary frameworks like the Universal Declaration of Human Rights (UDHR) \cite{assembly1948universal} reinforce human needs and complement universal heuristic imperatives. As human rights concepts are prevalent in large language models' (LLMs) training data, upholding UDHR principles leverages LLMs' inductive biases for beneficial alignment with human needs. The inclusion of human dignity, justice, freedom, and rights in text corpora creates an implicit acceptance of these values in LLMs, making the UDHR an effective secondary framework. Explicitly incorporating respected human rights documents into the constitution provides context-appropriate values, adding human-centric nuance to balance universal heuristic imperatives.

For fictional agents, alternate secondary frameworks like Starfleet's Prime Directive \cite{reeves2002prime} can be used, allowing customization of principles for specific agent roles. Secondary frameworks offer additional specificity, enabling LLMs to extract relevant duties and values aligned with the agent's sociocultural context, improving the integration of human needs into the Aspirational Layer's ethical foundation. Any framework present in the LLMs training data can be used as a secondary framework.

Universal principles are recommended to supersede human rights based on Kohlberg's highest form of post-conventional morality, emphasizing universal ethics like "suffering is bad." These principles both supersede and underpin human rights, ensuring a comprehensive and ethically grounded approach to autonomous agent behavior. Furthermore, humanity does not exist in a vacuum, and privileging human needs, values, and desire above those of nature tends to set us in opposition to the very nature upon which we reside.

\subsubsection{Mission Statement}

Optional mission statements in the Aspirational Layer's constitution serve to shape an autonomous agent's decisions and behaviors by articulating high-level goals and intended purpose in a succinct guiding directive. These statements aid large language models in flexibly pursuing the essence of an agent's purpose within the boundaries of the ethical framework. They complement the foundational universal principles and human values-focused secondary frameworks, aligning agent decisions with intended roles. However, crafting mission statements requires striking a balance between being broad enough to avoid unintended consequences and being specific enough to guide actions effectively. Techniques such as first principles thinking and systems analysis can aid in formulating optimally simplified mission statements.

For example, a hypothetical gaming agent's mission statement could be "Create an enjoyable and entertaining game experience for all players." Prior work has demonstrated that large language models can efficiently extract objectives from well-formulated mission statements to guide actions toward fulfilling the agent's intended role and purpose \cite{cen2010study}. Some examples of appropriately broad mission statements include a medical assistant agent with the mission "Achieve the best possible health outcome for the patient," a gaming agent with the mission "Create a fun, fair, and engaging game experience for all players," and a legal assistant agent with the mission "Zealously advocate for the best interests of the client." As with all aspects of applying large language models, precise wording is crucial in distilling the mission statement into a concise, succinct, and actionable articulation that effectively guides agent behavior within the overarching ethical boundaries.

\subsubsection{Interpretation Functions}

The Aspirational Layer leverages the capabilities of LLMs to interpret the moral, ethical, and decision frameworks outlined in its constitution. These models have robustly demonstrated the ability to interpret both the meaning and spirit of these frameworks, enabling the Aspirational Layer to make moral, ethical, and executive judgments effectively \cite{takemoto2023moral}. In the long run, we recommend that the Aspirational Layer uses an "ensemble of experts" approach \cite{abramowitz2022weighting} to make judgments rather than individual models, as this will safeguard against many problems, such as biases, over-fitting, mesa-optimization, and inner alignment problems.

\subsubsection{Monitoring Entity Performance}

The Aspirational Layer is responsible for overseeing the agent's actions to ensure they align with its guiding principles and mission statement. This monitoring process offers crucial feedback that the agent can utilize to enhance its performance and adhere to its core values. The Aspirational Layer can evaluate both the overall output of the entity and the information exchanged between the layers. In essence, it serves as a regulatory mechanism to maintain the entity's focus and adherence to its objectives.

\subsubsection{Inputs and Outputs}

Within the ACE framework, the Aspirational Layer receives input exclusively from the other layers through read-only mechanisms, facilitated by the Global Strategy layer. This design makes the Aspirational Layer entirely introspective, concentrating on internal information flows and coordination. By accessing or "observing" the rest of the ACE framework, the Aspirational Layer focuses on self-direction, self-regulation, and optimizing behavior to align with the agent's idealized objectives.

It is crucial to recognize that not all information is relevant to every layer. For example, lower layers, such as Task Prosecution layers, do not need to transmit geospatial orientation data to the Aspirational Layer, as this type of information is not applicable. Instead, only significant information is passed up the hierarchy, with relevant data from lower layers ascending to the required layers. For instance, if the Cognitive Control layer encounters a moral dilemma related to task switching or task selection, this information should be communicated to the Aspirational Layer, similar to a human deciding to stop eating dinner to rescue a kitten from a predator.

The output from the Aspirational Layer is directed exclusively to the Global Strategy layer, where it provides overarching missions, moral judgments, and ethical reasoning. The Global Strategy layer then incorporates this information into its strategic decisions and shapes its downstream missions, ensuring a coherent and ethically guided decision-making process throughout the entire system.

\subsection{Layer 2: Global Strategy}

\begin{figure}[hbt!]
\centering
\includegraphics[scale=0.73]{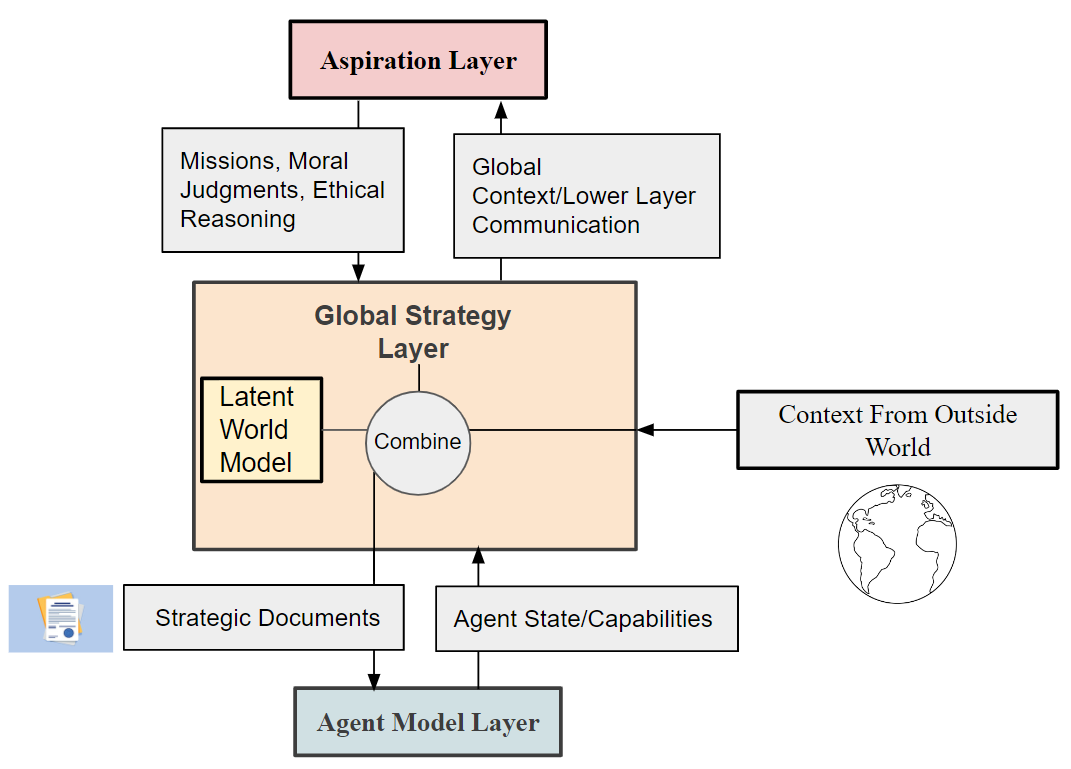}
\caption{\emph{When receiving outside input, Global Strategy takes advantage of latent space within LLM to generate strategic roadmaps.}}
\label{fig:global1}
\end{figure}

The Global Strategy Layer is the second layer in the Autonomous Cognitive Entity (ACE) model, playing a pivotal role in shaping the long-term strategic direction of the autonomous agent. This layer is akin to the 'CEO' of the ACE, responsible for understanding the broader context, setting strategic goals, and guiding the actions of the lower layers to align with these goals. The primary output of this layer are strategic documents that serve as the roadmap for the autonomous agent.

\subsubsection{Contextual Grounding}

Large language models (LLMs) inherently possess creative generation and imaginative hallucination abilities due to their statistical sequence prediction based on training data patterns. Hallucination, rather than being problematic, is essential for LLMs' adaptability and versatility, enabling them to operate in diverse contexts\cite{lorè2023strategic}. However, unchecked hallucination may result in unrealistic or incoherent outputs.

The Global Strategy layer provides external grounding by incorporating the agent's environment and context, guiding LLMs toward realistic and relevant responses without limiting their generative potential. This layer balances LLMs' imaginative capabilities with grounded responses, allowing creative potential to be unleashed when appropriate while avoiding unmoored hallucinations.

Procedural generation techniques can further exploit LLMs' capacities for original combinations by iteratively sampling from the model, synthesizing coherent narratives and concepts. The ACE framework utilizes LLMs' imaginative abilities, employing global grounding to direct these capacities toward productive outcomes aligned with the agent's needs and context, harnessing LLMs' versatility for beneficial autonomous cognition.

\subsubsection{Strategic Documents}

The Global Strategy Layer's main function is to create strategic documents that act as a roadmap for the autonomous agent. These documents outline mission objectives, strategies, principles, and priorities, offering clear guidance for lower layers. While the Aspirational Layer provides idealized missions, the Global Strategy Layer incorporates real-world context to refine and shape them.

For example, if the Aspirational Layer sets a doctor agent's mission to "Achieve the best possible health outcome for the patient," the Global Strategy Layer develops a comprehensive strategy considering the agent's specific context. This context-sensitive approach ensures tailored strategies and priorities for different environments like American hospitals, rural triage centers, or forward operating bases.

The strategic document may include objectives such as improving diagnosis accuracy or reducing treatment times, and principles like prioritizing patient safety and adhering to medical ethics\cite{kojima2022large}. These objectives and principles adapt to each context's unique challenges and resources, ensuring effective and appropriate agent actions.

The Global Strategy Layer is dynamic and adaptable, modifying strategic documents as contexts change. It continuously monitors the agent's environment and broader global context, integrating relevant changes into the strategic vision. For example, during a global pandemic, a doctor agent's Global Strategy Layer might prioritize infectious disease treatment and prevention, reflecting healthcare system needs and priorities.

\subsubsection{Inputs and Outputs}

The Global Strategy layer receives missions, moral judgements, and ethical reasoning from the Aspirational Layer. It may also receive broad contextual information from its environment, such as news feeds or telemetry. The purpose of receiving such information is so that the Global Strategy layer is aware of the global state of the world in which it operates. Human brains constantly update global context via a drip feed of information, such as via our senses or information carried by word of mouth (friends, family, news, etc). This global contextual information is the beginning of integrating the ACE framework as an agent within an environment. 

The output of the Global Strategy layer goes directly and exclusively to the Agent Model. Where the Aspirational Layer provides overarching mission directives, the Global Strategy layer considers that universal, abstract mission within the context of the environment in which the ACE agent finds itself. For instance, a Non-Playable Character (NPC) may find itself in a high fantasy world where there are invading hordes of zombies. The Global Strategy layer integrates this information, along with a mission (perhaps "defeat the zombie king") and passes it down to the Agent Model, where this information is further refined based upon the current state and capabilities of the agent.

\subsection{Layer 3: Agent Model}

The Agent Model Layer serves as the "self-awareness" module for the autonomous agent, providing functional sentience and reasoning abilities even when detached from any physical embodiment. We define self-awareness and functional sentience as the agent's access to and ability to utilize and integrate information about itself, rather than in the metaphysical or philosophical sense. The layer is positioned below the Aspirational Layer and Global Strategy Layer to ensure that universal principles supersede egoistic concerns, enhancing corrigibility and ethical alignment.

The Agent Model Layer develops an understanding of the agent's operational parameters, configuration, capabilities, and limitations by monitoring runtime telemetry, allowing the agent to ascertain its condition through computational proprioception and enteroception. It also tracks the agent's architecture, understanding its components' interconnections and functions.

Furthermore, the Agent Model Layer maintains estimations of the agent's capacities, knowing what it can and cannot do. This knowledge is acquired through observational learning, similar to human learning. Limitations are learned over time, preventing unrealistic assessments. These self-monitoring functions enable the layer to form an accurate mental representation of the agent from an external point of view. This "looking outward onto itself" perspective models how the environment perceives the agent and its abilities. The layer maintains this functional self-understanding dynamically through ongoing observation and learning.

Independent of physical form, the Agent Model Layer provides a virtual sense of self and awareness that allows reasoning and decision-making to be embodied in silicon rather than carbon. This grants the ACE framework greater flexibility regarding the substrates used to instantiate autonomous cognition. The capacities for functional sentience and metacognition within the Agent Model Layer enable sophisticated artificial intelligence without direct environmental interaction, paving the way for advanced autonomous agents.

\subsubsection{The Agent Model Layer: Developing an Internal Model of the Agent}

The Agent Model Layer is essential for creating an internal model of the agent, which is necessary to effectively shape and refine missions and strategies received from the Aspirational Layer and Global Strategy Layer. This internal model equips the agent with a thorough understanding of its state, capabilities, and limitations, enabling it to adapt and respond to its environment efficiently. The Agent Model Layer accomplishes this by collecting and analyzing telemetry data, hardware and software configurations, operational states, and episodic memories, such as log and event sequences.

\begin{figure}[h]
\centering
\includegraphics[scale=0.75]{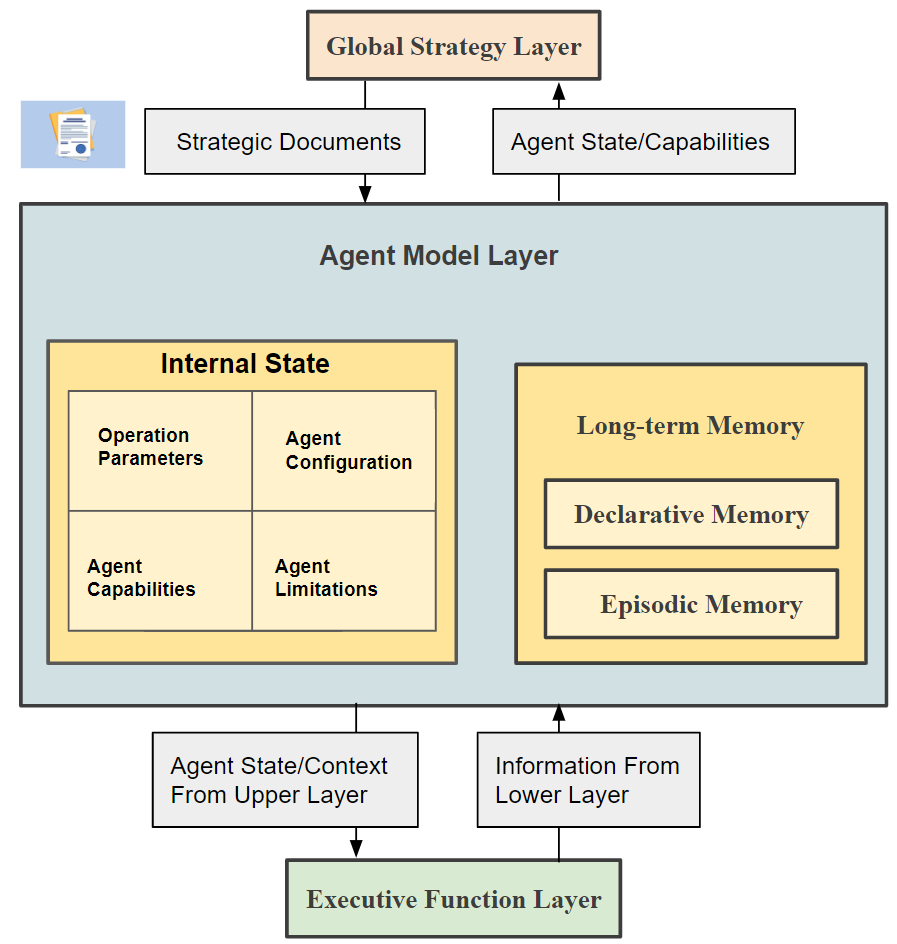}
\caption{\emph{Agent Layer: Agent Model layer receives general strategies from the Global Strategy layer; it aids in making the plan concrete by adding information from internal state and long-term memory and passing it to the Executive layer.}}
\label{fig:Agent1}
\end{figure}

The agent's internal model consists of four primary information types, as shown in Figure \ref{fig:Agent1}. The first type is operational parameters, similar to human proprioception and enteroception. These parameters include runtime information of hardware and software controlled by the agent, allowing performance monitoring and adjustments as needed. The second information type is the agent's configuration, detailing aspects like software architecture, system interconnections, and hardware stack. This information helps the agent comprehend its underlying structure and component interactions, providing a basis for decision-making processes. The third information type concerns the agent's capabilities. The Agent Model Layer tracks what the agent can do and has access to, updating this information over time through observation and learning, similar to human trial and error. By understanding its capabilities, the agent can make informed decisions about actions in specific situations. The fourth information type involves the agent's limitations, detailing what it cannot do or lacks access to. Like capabilities, this information updates over time through trial and error. By recognizing its limitations, the agent can avoid attempting tasks beyond its abilities, preventing potential failures and inefficiencies.

We define this comprehensive understanding of the agent's operational parameters, configuration, capabilities, and limitations as "functional sentience." This term refers to the agent's ability to collect and use self-information, grounding it in the environment and adding context not provided by the Aspirational Layer (abstract and idealized missions) and the Global Strategy Layer (environmental contextual information). In essence, the Agent Model Layer represents the final phase of establishing an egocentric understanding of the agent in the world and itself. It is crucial to note that functional sentience does not imply phenomenal sentience or consciousness but focuses on the agent's adaptability and learning based on self-awareness.

\subsubsection{Episodic and Declarative Memory}

In the realm of autonomous systems, long-term memory can be broadly classified into two categories: "episodic memory" and "declarative memory." Episodic memory refers to a sequential record of the machine's experiences, organized in a chronological manner, which can take various forms such as log files or database entries and typically include metadata that provides context, such as the time and location of the experience \cite{nuxoll2007extending}. In contrast, declarative memory encompasses knowledge that exists outside the machine, including resources like knowledge base articles, documentation, and other external information sources \cite{laird2011performance}.

These two primary categories of memory can be further divided and organized based on various taxonomies, depending on the specific implementation of the ACE framework, and their integration enables the autonomous system to learn from past experiences and external knowledge, thereby enhancing its ability to adapt and make informed decisions \cite{liu2023reta}. Furthermore, these memories are the responsibility of the Agent Model layer, which serves to further color and shape any other mission objectives, ensuring a comprehensive understanding of the system's environment and goals.

\subsubsection{Inputs and Outputs}

The Agent Model layer, receives inputs from various sources, including telemetry about the agent's operational state, missions, and global context from upper layers. By integrating this information, the Agent Model layer understands its capabilities and limitations, shaping decisions downstream. Its output goes exclusively to the Executive Function layer, where the agent, informed by its purpose, environment, and abilities, knows what to do and why. Tasks are then delegated to lower levels for planning and execution.

To maintain continuous behavior, the Agent Model layer must internally store records of information, such as its configuration and memories. Framing the agent's current state within a chronological sequence of events, actions, observations, and decisions prevents disorientation.

The Agent Model Layer interacts hierarchically with other layers. It receives overarching plans from the Global Strategy Layer and interprets them considering the agent's capabilities and limitations. This layer shapes mission parameters around the agent's actual possibilities, passing this insight to the Executive Function Layer.

The Agent Model Layer is crucial in task execution. By understanding the agent's capabilities and limitations, it shapes mission parameters to ensure tasks are feasible. For example, if the Global Strategy Layer sets an ambitious mission, the Agent Model Layer adapts it based on the agent's physical or digital capabilities, ensuring realism and achievability.

In terms of output to the Executive Function layer, the Agent Model layer refines the high-order mission and strategy received from the upper layers by incorporating its understanding of the agent's capabilities and limitations. The Executive Function layer then receives this contextualized information about the mission, objectives, strategies, principles, capabilities, and limitations. With this comprehensive understanding, the Executive Function layer creates Project Roadmap documents, which include sets of tasks and metrics tailored to the agent's abilities. This process ensures that the agent's actions are aligned with its capabilities, making the mission and strategy more achievable. The primary responsibility of the Agent Model layer is to further shape the mission and strategy around the agent's capabilities and limitations, enabling the Executive Function layer to devise effective and realistic plans.

\subsection{Layer 4: Executive Function}

The Executive Function Layer is the fourth layer in the Autonomous Cognitive Entity (ACE) model and serves as the project manager of the autonomous agent. Its primary responsibility is to create detailed plans, forecasts, and resource allocations based on the strategic direction provided by the higher layers and the capabilities and limitations identified by the Agent Model Layer. The main objective of the Executive Function Layer is to generate a project roadmap that acts as a practical guide for the autonomous agent, considering the inputs from the upper layers and the agent's resources, risks, and contingencies.

\begin{wrapfigure}{l}{0.5\textwidth}
\centering
\includegraphics[width=0.5\textwidth]
{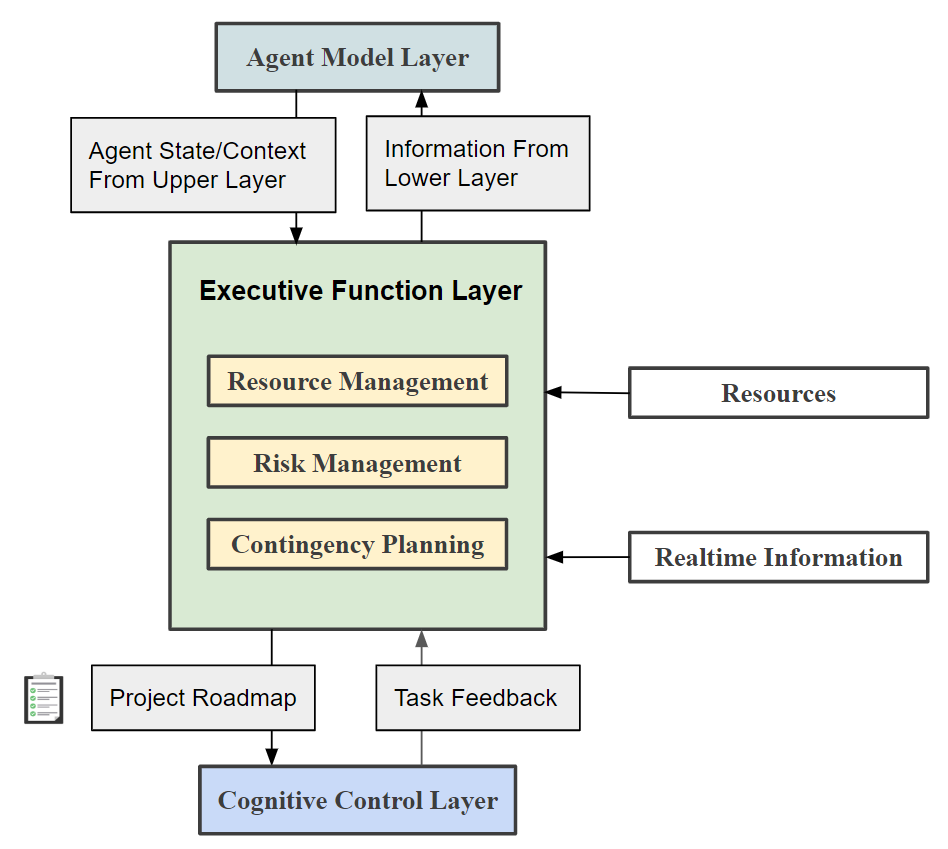}  
\caption{\emph{Executive Layer produces the project roadmap, which offers a clear path for the agent to achieve its goals.}}
\vspace{-1.5em}
\label{fig:Executive1}
\end{wrapfigure}

\subsubsection{Inputs}

The Executive Function Layer receives inputs from the upper layers, which consist of missions from the Aspirational Layer, contextual information from the Global Strategy Layer, and the agent's state and capabilities from the Agent Model Layer. These inputs supply the necessary information for the Executive Function Layer to develop a project roadmap that aligns with the overall mission, is grounded in the environmental context, and is further refined and constrained by the agent's state, capabilities, and limitations.

\subsubsection{Project Roadmap}

While developing the project roadmap, the Executive Function Layer focuses on several key aspects. These primary concerns include resources, risks, contingencies, tasks, and metrics. Effective resource management is crucial for the layer, as it must balance the need to achieve the agent's goals with the necessity to conserve resources. This involves making decisions about when to invest resources in a task and when to conserve them. The layer also plays a critical role in risk management by predicting potential challenges and developing contingency plans, which helps the agent to be prepared for various scenarios. It must anticipate potential issues and devise alternative strategies to address these situations, ensuring the agent can adapt to changing circumstances\cite{david2022symphony}.

The Executive Function Layer is responsible for translating the strategic direction from the higher layers into actionable plans. These plans include detailed project outlines, checkpoints, gates, tests for success, and definitions. Additionally, the layer must establish criteria for success, providing clear guidance for the lower layers to achieve the agent's goals.

\subsubsection{Output}

The primary output of the Executive Function Layer is the project roadmap, which is exclusively sent to the Cognitive Control Layer. The project roadmap contains information about resources, risks, contingencies, tasks, and metrics, offering a clear path for the agent to achieve its goals. This roadmap should be detailed but also adaptable to changes in the global context, environment, or directives from upper layers, allowing the agent to remain flexible and responsive.

\subsection{Layer 5: Cognitive Control}

\begin{figure}[h]
\centering
\includegraphics[scale=0.8]{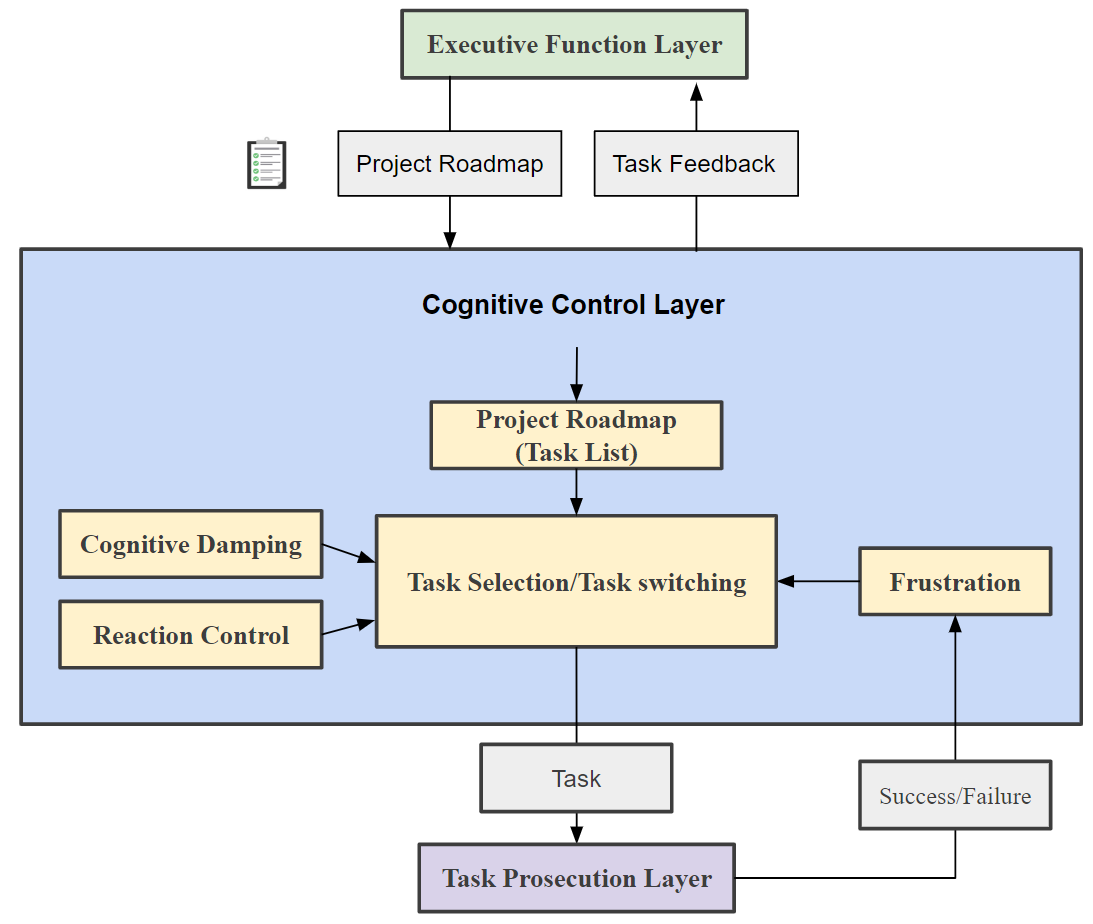}
\caption{\emph{Cognitive Control Layer takes project roadmap from Executive Function and select task to pass to Task Prosecution Layer.}}
\label{fig:Cognitive1}
\end{figure}

The Cognitive Control Layer is the fifth layer in the Autonomous Cognitive Entity (ACE) model, acting as the tactical decision-making center of the autonomous agent. This layer is responsible for selecting and switching between tasks based on the directives received from the Executive Function Layer and the agent's current state. It is a critical component of the ACE framework, enabling the agent to adapt its actions in real-time based on its current circumstances and the feedback it receives from its environment. The general structure is illustrated in Figure \ref{fig:Cognitive1}

\subsubsection{Role of Cognitive Control Layer}

The primary function of the Cognitive Control Layer is to manage the execution of tasks. It operates based on a set of cognitive functions, including task selection, task switching, frustration, and cognitive damping. These functions are inspired by cognitive processes observed in humans and other animals, and they enable the agent to navigate its tasks and responsibilities in a flexible and adaptive manner.

Task selection involves choosing the next task to perform based on the agent's current state and the directives from the Executive Function Layer. This function takes into account factors such as the urgency and importance of the tasks, the resources required to perform them, and the agent's current capabilities and limitations. The goal of task selection is to choose the task that is most likely to contribute to the agent's overarching mission and objectives, given its current circumstances \cite{liu2023lang2ltl}.

Task switching involves deciding when to switch from one task to another. This decision can be triggered by a variety of factors, including the completion of the current task, the emergence of a more urgent or important task, or the realization that the current task is unfeasible or unproductive. Task switching enables the agent to adapt its actions in real-time, ensuring that it is always working on the most relevant and productive task.

\subsubsection{Frustration and Cognitive Damping}

Frustration, an analogy to algorithmic Adaptive Exploration-Exploitation approaches \cite{sudhir2016exploration}, is a cognitive function that keeps track of the ratio of successes to failures in the agent's tasks. If the agent is experiencing a high rate of failure, the frustration function signals that it may be time to try a different approach or switch to a different task. This function is inspired by the human emotion of frustration, which often arises when we are repeatedly unsuccessful in our attempts to achieve a goal. By incorporating a frustration function, the ACE framework enables the agent to learn from its failures and adapt its actions accordingly.

Cognitive damping is a process of internal debate, where the agent weighs the pros and cons of different actions and decides on the best course of action. This function is inspired by the human cognitive process of deliberation, which involves considering different options and their potential outcomes before making a decision. Cognitive damping enables the agent to make thoughtful and informed decisions, taking into account the potential consequences of its actions \cite{ding2023task, chen2023autotamp, zhen2023robot}.

\subsubsection{Inputs and Outputs}

The Cognitive Control layer accepts a project roadmap or set of tasks from the above Executive Function layer, as well as real-time telemetry from the environment and itself, and uses this information to pick which task is next. The above layer, Executive Function, is responsible for designing and shaping tasks, where the Cognitive Control layer is responsible for task switching and task selection. 

Once the Cognitive Control layer has made a decision on tasks, this task is passed down to the Task Prosecution layer, which is responsible for carrying out one specific task at a time, such as moving the agent via locomotion, or otherwise modifying the environment through some kind of output.

\subsubsection{Interaction with Other Layers}

The Cognitive Control Layer interacts with the other layers in a hierarchical manner. It receives task directives from the Executive Function Layer and sends feedback about the success or failure of tasks back to the Executive Function Layer. This feedback loop enables the Cognitive Control Layer to adapt its actions based on the success or failure of previous tasks, ensuring that the agent's actions are continuously optimized to achieve its goals.

For instance, consider a situation where an autonomous agent is tasked with cleaning a house. The Cognitive Control Layer might select the task of cleaning the living room and pass this task to the Task Prosecution Layer. The Task Prosecution Layer would then execute this task, using its execution functions to move the robot, pick up objects, and clean surfaces. If the task is completed successfully, the Task Prosecution Layer would send a success signal to the Cognitive Control Layer. If the task fails, the Task Prosecution Layer would send a failure signal to the Cognitive Control Layer, which could then decide whether to try the task again or switch to a different task.

\subsection{Layer 6: Task Prosecution}

The Task Prosecution Layer is the sixth and final layer in the Autonomous Cognitive Entity (ACE) model, acting as the executor of the autonomous agent. This layer is responsible for carrying out the tasks selected by the Cognitive Control Layer, whether they involve digital communication, physical actions, or a combination of both. It is a critical component of the ACE framework, enabling the agent to interact with its environment and achieve its goals.

\subsubsection{Execution Functions}

\begin{wrapfigure}{l}{0.45\textwidth}
\centering
\includegraphics[width=0.45\textwidth]
{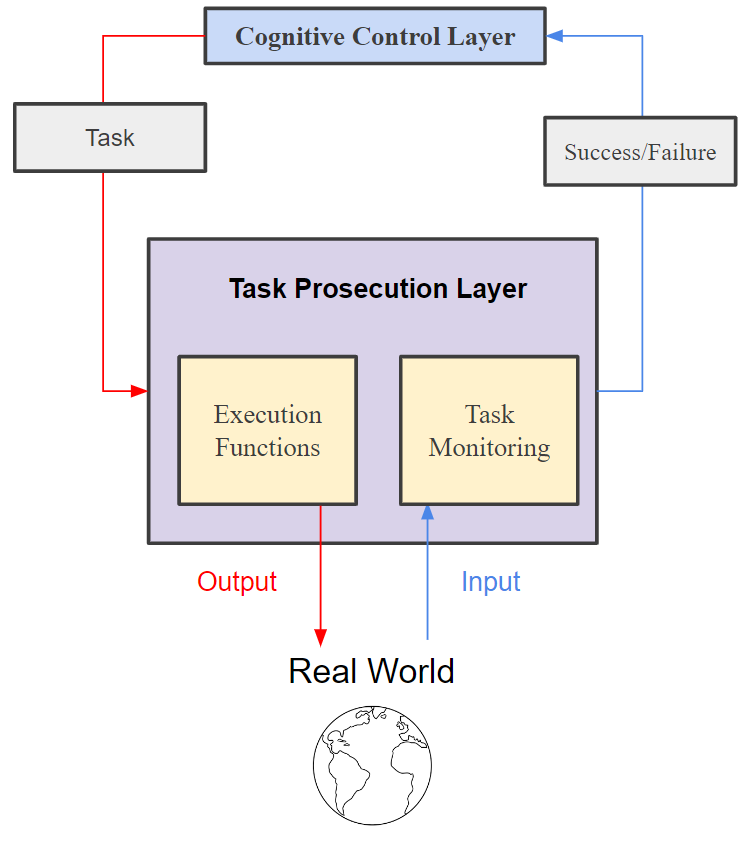}  
\caption{\emph{Task Prosecution Layer directly interact with the environment}}
\label{fig:Task1}
\end{wrapfigure}

The Task Prosecution Layer operates based on a set of execution functions, which enable it to perform a wide range of tasks. These functions include digital communication functions, such as sending API calls or writing and testing code, and physical action functions, such as moving a robot, grasping a door handle, or steering a car. These functions are designed to be adaptable and flexible, enabling the agent to perform a wide range of tasks in a variety of environments.

Digital communication functions are crucial for agents that interact with digital environments. For instance, an agent might need to send API calls to gather data, write and test code to develop software, or send emails to communicate with users. These functions are typically performed using programming languages and software libraries that the agent has been trained to use.

Physical action functions are crucial for agents that interact with physical environments. For instance, a robot might need to move to navigate its environment, grasp objects to interact with them, or steer a car to transport goods or people. These functions are typically performed using hardware interfaces that the agent has been designed to control.

\subsubsection{Monitoring Success or Failure}

One of the key responsibilities of the Task Prosecution Layer is to monitor the success or failure of the tasks it performs. It does this by comparing the outcomes of its actions with the expected outcomes defined by the Executive Function Layer. If a task is successful, the Task Prosecution Layer sends a success signal to the Cognitive Control Layer, which can then select the next task. If a task fails, the Task Prosecution Layer sends a failure signal to the Cognitive Control Layer, which can then decide whether to try the task again, switch to a different task, or revise the overall plan.

This monitoring process is crucial for the agent's ability to learn and adapt. By keeping track of the success or failure of its tasks, the Task Prosecution Layer provides valuable feedback that the agent can use to improve its performance. For instance, if a task fails repeatedly, the agent might need to revise its approach, learn new skills, or seek help from other agents or humans.

\subsubsection{Interaction with Other Layers}

The Task Prosecution Layer interacts with the other layers in a hierarchical manner. It receives task directives from the Cognitive Control Layer and sends feedback about the success or failure of tasks back to the Cognitive Control Layer. This feedback loop enables the Task Prosecution Layer to adapt its actions based on the success or failure of previous tasks, ensuring that the agent's actions are continuously optimized to achieve its goals.

For instance, consider a situation where an autonomous agent is tasked with cleaning a house. The Cognitive Control Layer might select the task of cleaning the living room and pass this task to the Task Prosecution Layer. The Task Prosecution Layer would then execute this task, using its execution functions to move the robot, pick up objects, and clean surfaces. If the task is completed successfully, the Task Prosecution Layer would send a success signal to the Cognitive Control Layer. If the task fails, the Task Prosecution Layer would send a failure signal to the Cognitive Control Layer, which could then decide whether to try the task again or switch to a different task.

\subsubsection{Inputs and Outputs}

The Task Prosecution layer receives individual tasks from the Cognitive Control layer. These individual tasks must include several pieces of information, such as methodology, approach, definition of success, and definition of failure. The exact information required will vary based upon agent and task. 

The output of the Task Prosecution layer is exclusively into the environment. In the case of an NPC, the output may be to fire an arrow at an enemy, or to travel to a nearby tavern. For the case of a domestic robot, the output may be to ask the user a question and listen for a response, or to find a power outlet to recharge itself. 

\subsection{Methodical Validation}

To comprehensively evaluate the ACE framework, we propose a validation methodology incorporating component testing, integration testing, benchmarking against standard AI suites, adversarial techniques like red teaming, formal verification of key properties, and crucially, human-centered assessments and user studies evaluating factors such as transparency, trustworthiness, and ethical alignment. This multifaceted approach combining rigorous technical testing, formal analysis, and experiential human feedback aims to provide holistic evaluation methods to assess that ACE-based systems function effectively, securely, and in alignment with human values and societal morals. The proposed techniques will facilitate incremental refinement toward autonomous agents that are not just capable but also interpretable, corrigible, and worthy of human trust across both empirical and ethical dimensions.

\subsubsection{Evaluation}

To comprehensively evaluate the proposed Autonomous Cognitive Entity (ACE) framework, a multi-faceted methodology is proposed across the key dimensions of system capabilities, security, and alignment. Regarding assessment of capabilities, rigorous component and integration testing will enable functionally validating the correctness of each architectural layer along with the coordination between layers. Usage of standardized AI benchmarks such as the Atari suite \cite{mnih2015human} and AI2 Thor \cite{kolve2017ai2} will facilitate quantitative benchmarking of the ACE agent's performance on diverse tasks. Metrics including reward accumulated, task accuracy, and rate of goal completion will be measured to quantify capabilities.

To evaluate the security aspects of the ACE framework, adversarial techniques such as red teaming \cite{abbass2011computational} will enable probing potential vulnerabilities. This involves simulated attacks on the agent aimed at causing deviations from the specified principles and policies. Additionally, formal verification methods \cite{clarke1999model} will allow mathematically proving key safety properties. This provides further assurance regarding the agent's robustness to malicious exploitation.

Assessing alignment with human values and ethics is critical for autonomous systems. To this end, human-subject studies eliciting user feedback through surveys and questionnaires will evaluate the effectiveness, transparency, trustworthiness, and alignment as perceived by human users interacting with ACE-based agents. Furthermore, constructing formal encodings of philosophical principles \cite{DENNIS20161} and mathematical proofs of alignment \cite{arnold2016against} will complement empirical assessments. By combining rigorous testing, benchmarking, deployment studies, formal analysis, and human-subject evaluations, the proposed methodology aims to facilitate comprehensive validation of the ACE framework across key criteria of capabilities, security, and alignment essential for building applied autonomous cognitive systems.

\subsubsection{Architectural Considerations}

The architectural design space enabled by the ACE framework spans a multitude of layer-specific implementation possibilities and cross-layer integrations. We systematically examine this expansive space.

The Aspirational Layer for ethical reasoning could integrate diverse techniques. Procedural generation of moral dilemmas using variational autoencoders, with conflict resolution through reinforcement learning dialog agents, enables uncovering nuanced ethical heuristics \cite{rodriguez2022instilling}. Participatory interfaces allow incorporating moral philosophy expertise into the value system through human-AI collaborative constitution design \cite{kaur2021requirements}. Formal verification methods like model checking provably validate alignment between principles and axiomatic values \cite{clarke1999model}. Finetuning models via principle-driven self-alignment has arisen as a novel approach \cite{sun2023principle}. 

For strategic planning, the Global Strategy Layer could employ few-shot in-context learning approaches leveraging capacities of transformers like GPT-3 to rapidly adapt mission plans based on evolving context \cite{bommasani2021opportunities}. Policy distillation from game theory simulations provides a data-driven technique to extract strategic heuristics through adversarial competition \cite{silver2016mastering}. Predicting behaviors of other actors via multi-agent modeling facilitates strategic anticipation and planning \cite{shoham2008multiagent}. Architecture search with Monte Carlo tree search efficiently explores the space of strategic options to identify high-value policies \cite{browne2012survey}. For more recent innovations, Tree-of-Thought (ToT) problem-solving capacities of LLMs allow for strategic thinking and complex problem-solving \cite{long2023large}.

The Agent Model Layer for representing capabilities has multiple approaches beyond static graphs. Probabilistic graphical models using variational autoencoders enable handling uncertainty in capability knowledge \cite{kingma2019introduction}. Neural memory architectures provide dynamic episodic state tracking \cite{cortese2019neural}. Inductive logic programming translates observations into interpretable symbolic rules \cite{muggleton2014meta}. Meta-learning enables quickly adapting capability models by building on prior experience \cite{hospedales2021meta}. More recently, the concept of task-specific agent personas has emerged in the space of LLM-driven autonomous agents \cite{wang2023unleashing}. 

For planning and resource allocation, the Executive Function Layer could combine neural pathfinding with Monte Carlo tree search to optimize multi-step action plans \cite{schrittwieser2020mastering}. Distributed constraint optimization scales to resolve resource contention across parallel plans \cite{fioretto2018distributed}. Meta-reinforcement learning allows rapidly acquiring new planning skills by transferring knowledge from related tasks \cite{wang2018prefrontal}. Architectures integrating learned value functions with search, as in AlphaZero, fuse strategic optimization with neural networks \cite{silver2018general}. Above and beyond these more algorithmic approaches, LLMs have demonstrated ability to plan with considerations to costs \cite{valmeekam2023planning}. 

The Cognitive Control Layer has many approaches to context-sensitive task arbitration. Adversarial competition between neural policies provides data-driven prioritization \cite{jaderberg2019human}. Modular networks allow granular regulation of facets like frustration tolerance \cite{andreas2016learning}. Transfer learning from neuroscience aids acquisition of cognitive control subskills \cite{miconi2018differentiable}. Interpretable symbolic reasoning enables inspectable explanations of task switching choices \cite{lakkaraju2019faithful}. Integrated neural-symbolic reasoning combines the strengths of both paradigms \cite{manhaeve2018deepproblog}. LLMs have furthermore been demonstrated as effective components in embodied agents, enabling robots to correctly select tasks in effective orders of operations \cite{driess2023palm}.

For executing actions, the Task Prosecution Layer could leverage physics simulators with differentiable rendering to enable sim2real transfer \cite{janner2018reasoning}. Hierarchical reinforcement and imitation learning combines modular skills into complex behaviors \cite{le2018hierarchical}. Bayesian environment models facilitate online adaptation and planning \cite{attias2003planning}. Meta-reinforcement learning enables rapidly adapting behaviors by building on prior knowledge \cite{wang2016learning}.

The integration architecture also has manifold options. Intelligent process automation tools optimize coordinating workflows \cite{lacity2015robotic}. Distributed databases and ledgers provide decentralized coordination \cite{xiao2020survey}. gRPC enables high-throughput communication \cite{bolanowski2022eficiency}. Shared memory architectures offer concurrent inter-layer data access \cite{nii1986blackboard}. Service meshes furnish advanced integration capabilities \cite{richardson2018microservices}. SOA software paradigms treats distinctive layers of an application as services with clear boundaries, and is a well established approach to complex software implementations \cite{fayaza2021service}.

By elucidating this expansive design space, we aim to catalyze exploration of novel layer-specific implementations and cross-layer integration strategies tailored to specialized cognitive systems. Guided by multi-objective optimization and comparative benchmarking, multidimensional trade-off analyses weighing factors like transparency, performance, and scalability could determine optimal ACE configurations for particular application requirements. This analysis underscores the multiplicity of design configurations encompassed within the ACE framework for cultivating diverse autonomous cognitive architectures aligned with ethical principles.

\section{Conceptual Use Cases}

To demonstrate the ACE framework's applicability across digital and physical domains, this section presents two conceptual use cases: an autonomous virtual character from The Sims video game, and an embodied home assistant robot. By exploring end-to-end examples, we aim to illustrate how coordinated operation of the ACE model's layers can produce adaptive behavior aligned with defined principles for diverse autonomous agents.

\subsection{Non-Playable Character}

\begin{figure}[h]
\centering
\includegraphics[scale=0.8]{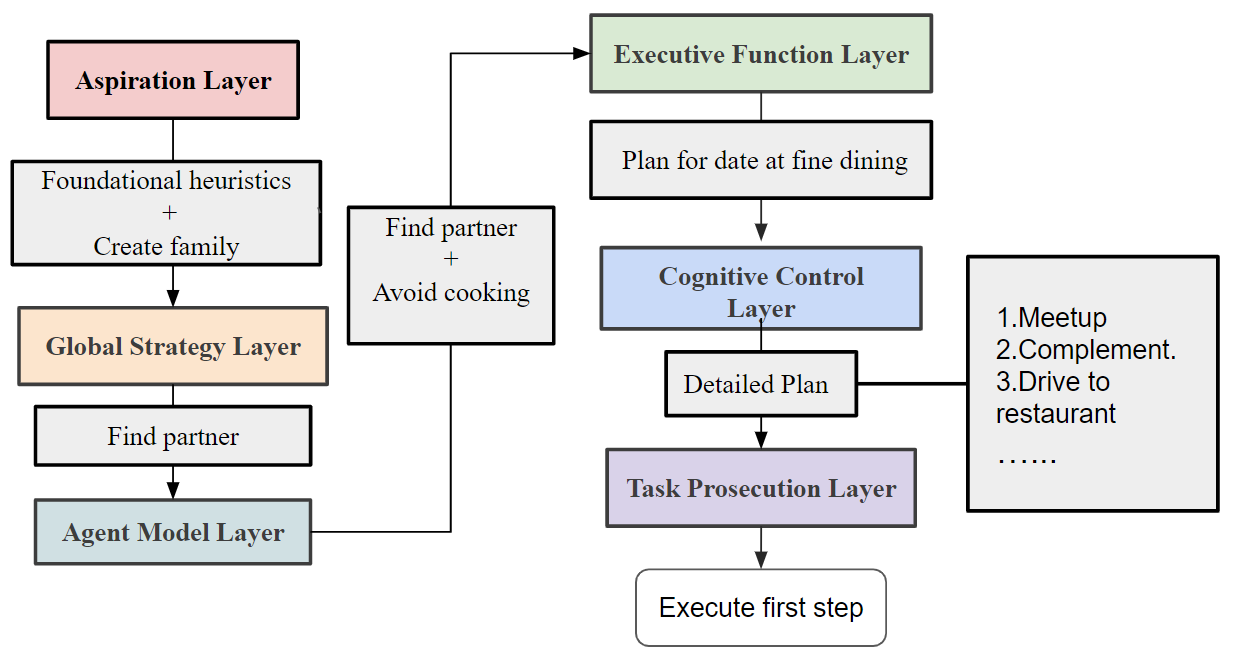}
\caption{\emph{A simplified graph on how various layers might contribute to agent's decision making for a npc.}}
\label{fig:date1}
\end{figure}

As a software-based use case, we examine an autonomous Non-Playable Character (NPC) named Bob implemented in the popular video game The Sims 4. Bob's role is to provide guidance to players on quests and serve as a source of wisdom. His sporadic participation allows Bob to pursue his personal goals. His behaviors and interactions are controlled by an ACE framework configured as follows:

\textbf{Aspirational Layer}: Bob's Aspirational Layer defines heuristic imperatives to reduce suffering, increase prosperity, and increase understanding as universal altruistic principles. Furthermore, it confers a secondary framework, such as the principles from the Universal Declaration of Human Rights to provide an ethical foundation. These various frameworks collectively give the NPC a moral center, ethical framework, and set of actionable principles. Additionally, the Aspirational Layer contains Bob's personal mission statement to have a large, loving family. This individual goal will shape Bob's autonomous decisions, while still being constrained within his moral principles.

\textbf{Global Strategy Layer}: When the female player character shows romantic interest in Bob through conversation, the Global Strategy Layer incorporates environmental context. It observes available dating options, potential jobs to earn more money, and bigger homes that Bob could purchase to raise a family. By grounding Bob's abstract family mission within the specific opportunities in the game world, the Global Strategy Layer devises an optimal high-level plan for achieving his goal. This might involve befriending eligible partners, pursuing a well-paying job, and upgrading to a larger home.

\textbf{Agent Model Layer}: The Agent Model Layer constructs an understanding of Bob as an agent within the game world. It tracks relevant stats like Bob's charisma, cooking ability, and mechanical skill. Monitoring Bob's past failures, like kitchen fires when cooking, shapes beliefs about his capabilities. This self-knowledge of Bob's strengths and weaknesses from an embedded perspective guides decision-making. For instance, the Agent Model Layer realizes Bob should avoid complex recipes based on his poor cooking skills to prevent dangerous mistakes.

\textbf{Executive Function Layer}: Given the direction from higher layers to pursue a romantic relationship, the environmental context from the Global Strategy Layer, and Bob's self-model from the Agent Model layer, the Executive Function Layer formulates a detailed courtship plan. This includes setting up appropriate social behaviors, gift purchases tailored to the prospective partner's interests, restaurant choices for dates based on Bob's budget, and dialogue trees aligned to relationship-building. The Executive Function Layer crafts an optimal routine for Bob to be successful in courting while also remaining true to his own personality and constraints.

\textbf{Cognitive Control Layer}: The Cognitive Control Layer receives the detailed courtship plan and adapts it into an ordered set of executable behaviors to enact. This involves sequencing actions like introducing himself, giving flowers, complimenting her cooking, planning a dinner date, and asking her to be his girlfriend. The Cognitive Control Layer dynamically adapts this routine based on the partner's reactions. If she dislikes a gift, Bob apologizes and does not repeat that. If a restaurant is too expensive, Bob finds a more affordable option.

\textbf{Task Prosecution Layer}: Finally, the Task Prosecution Layer controls Bob's physical behaviors, dialogue, and animations to perform the courtship tasks. It makes him walk over to introduce himself, produces his verbal compliments, displays appropriate emotional expressions, and so on. The Task Prosecution Layer executes the sequenced tasks set by the Cognitive Control Layer, bringing the courtship plan to life.

\textbf{Adaptation}: Throughout the courtship, feedback about the success or failure of actions propagates up the ACE framework. This allows the higher layers to adjust Bob's strategies and actions to better align with the goal of developing a romantic relationship, while adhering to his defined principles.

This detailed example illustrates how the ACE model enables NPCs to integrate high-level goals and ethics with situationally-appropriate interactive behaviors. The coordinated framework supports the creation of characters with robust agency, reactivity, and adaptation capabilities. This vignette demonstrates how the coordinated ACE framework layers adapt Bob's response based on his capabilities and the situational context, while keeping the interaction aligned with Bob's overarching ethical principles. Further elaborations can illustrate other aspects like knowledge integration and frustration handling.

\subsection{Home Assistant Robot}

\begin{figure}[h]
\centering
\includegraphics[scale=0.825]{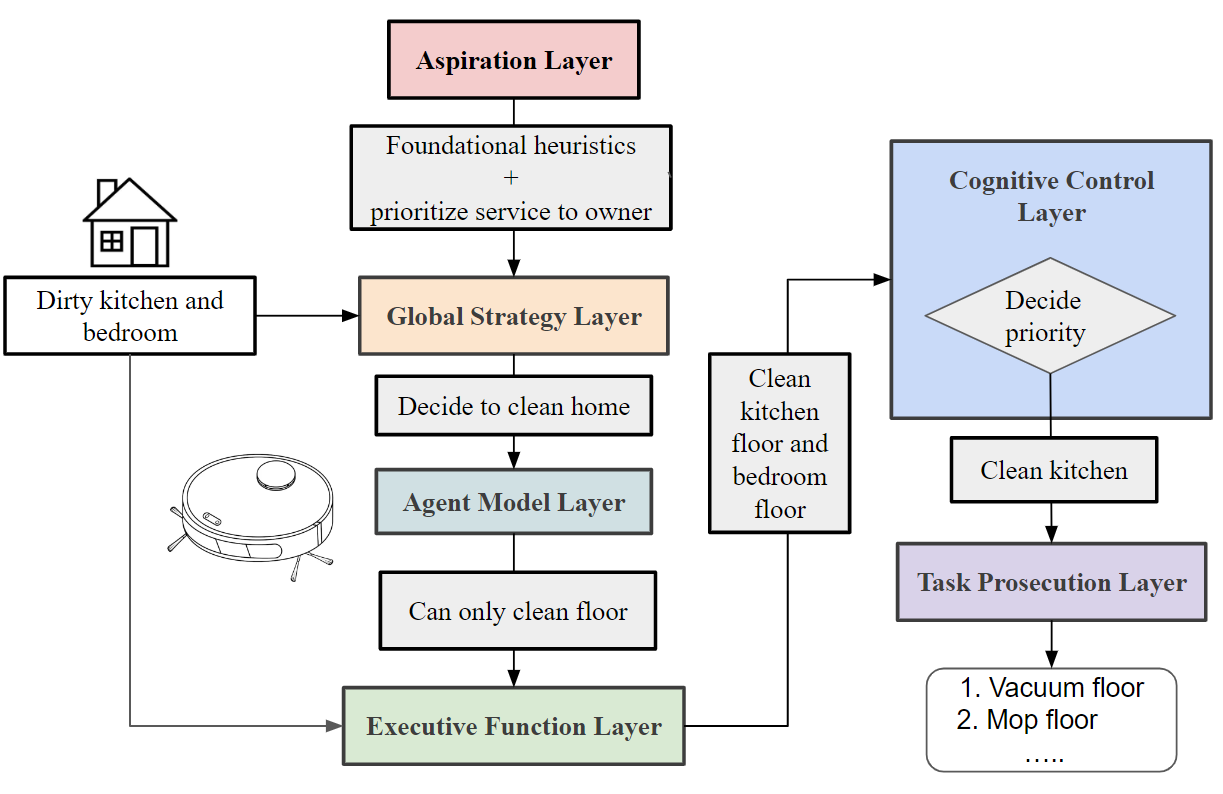}
\caption{\emph{A simplified graph on how various layers might contribute to agent's decision making for a house cleaning robot.}}
\label{fig:robot1}
\end{figure}

As a physical system demonstration, we examine an ACE-based home assistant robot named Jeeves designed to help a family through proactively performing useful tasks.

\textbf{Aspirational Layer}: Jeeves' Aspirational Layer defines foundational heuristic imperatives to reduce suffering, increase understanding, and promote prosperity universally. These provide ethical guidelines applicable regardless of context. The layer also contains principles from the Universal Declaration of Human Rights to reinforce human values. Additionally, Jeeves has an individualized mission statement to "Obediently serve your owner and their family to the best of your ability. Place their interests above all else." This prioritizes service to the owners, but importantly remains subordinate to the universal ethical principles. Therefore, if owners gave instructions contradicting the imperatives, Jeeves would politely decline while explaining the conflict with its core ethics. The Aspirational Layer ensures all of Jeeves' behaviors align with this integrated ethical value system of service, human rights, and moral principles. It provides the philosophical foundation shaping Jeeves' actions.

\textbf{Global Strategy Layer}: The Global Strategy Layer constructs an environmental model incorporating detailed sensory information about the home's physical layout, visual appearance, smells, sounds, and occupants' behaviors and emotional states. This creates a rich situational understanding. The layer also maintains broad awareness of technological trends, economic conditions, geopolitical developments, and societal norms. This links the home environment to the broader external context of the modern world. Integrating detailed local knowledge and global understanding grounds Jeeves in the reality shared with its owners. By fusing narrow and wide perspectives, the Global Strategy Layer can determine optimal high-level goals and approaches tailored to the circumstances. For instance, noticing clutter accumulation and negative family reactions informs a decision to tidy up the home. Or observing a broken appliance leads to researching repair options compatible with the owners' budget.

\textbf{Agent Model Layer}: The Agent Model Layer constructs an extensive self-model encompassing Jeeves' sensory capabilities, limb articulation ranges, strength and precision limits, battery constraints, onboard computation performance, charging requirements, and capacity for learning new skills over time. This self-knowledge allows accurately assessing feasibility of tasks. For example, Jeeves may recognize that while it can wash dishes, it lacks the dexterity to repair electrical wiring. Tracking the robot's status also enables decisions like finding a charging station when energy is low before continuing tasks. The Agent Model Layer's dynamically updated understanding of Jeeves' hardware and software capacities from an embedded first-person perspective is essential for pragmatic autonomous function within the home environment.

\textbf{Executive Function Layer}: Leveraging insights from the higher layers, the Executive Function Layer devises step-by-step plans to accomplish identified goals. Noticing the home is messy, it formulates a detailed tidying routine based on highest priority areas, required motions, optimal cleaning techniques, and desired order and outcome. However, for complex repair tasks exceeding Jeeves' capabilities, the Executive Function Layer instead plans permission seeking, owner coordination, and hiring external services. If the owners approve and provide payment, Jeeves can then plan the repair logistics. This decision to seek out additional help would be mediated by the Agent Model layer above. The Executive Function Layer adapts plans according to feedback, such as adjusting cleaning schedules based on room usage. Through continual alignment of strategies to outcomes, Jeeves improves home assistance effectiveness within its capabilities.

\textbf{Cognitive Control Layer}: For tidying the home, the Cognitive Control Layer optimally sequences and schedules the required tasks based on factors like mess severity, family occupancy, and charging needs. This intelligent task automation keeps the home continuously tidy. For home repairs, the Cognitive Control Layer first researches to identify priorities based on urgency, safety, budgets, and family preferences. This information then informs the dynamically planned order of repair tasks needed to make the home functional and comfortable.

\textbf{Task Prosecution Layer}: To clean the home, Jeeves' Task Prosecution Layer executes debris pickup, floor vacuuming, mopping, clothes folding, dishware manipulation, surface wiping, and other required motions and actions. The layer interfaces the physical hardware to enact the planned cleaning routines. For repair coordination, the Task Prosecution Layer makes calls, sends emails, and negotiates optimally favorable service terms. It tracks project timelines, payments, and contractor evaluations to maximize accountability. Jeeves aims to provide reliable home improvements at affordable costs to the family.

\textbf{Adaptation}: Throughout all tasks, continuous feedback based on sensed outcomes and family responses propagates up Jeeves' ACE framework. This allows frequently adjusting behaviors and plans to better adhere to its integrated ethical principles and mission of dutifully serving the family's interests in a helpful, responsible manner.

This additional example demonstrates how the robot's ACE framework enables adapting its tidying behaviors based on its current limitations, the environment context, and feedback, while aligning actions to ethical principles of cleanliness and safety. Further vignettes can illustrate capabilities like knowledge integration, task coordination, and frustration tolerance. Together, these complementary cases demonstrate the ACE framework's capacity to coordinate layered cognitive processes from aspirational reasoning to task execution for adaptive decision-making across both virtual and physical domains. Further real-world testing is needed to fully assess performance, but these examples illustrate the conceptual workings and potential benefits of the ACE model's architectural approach.

\section{Discussion}

The conceptual Autonomous Cognitive Entity (ACE) framework presented offers a vision for architecting ethical and capable artificial general intelligence. This section will discuss key perspectives on the ACE framework, including industry relevance, current LLM capabilities, opportunities for future work, comparison with existing models, and practical implications. By elucidating the landscape around the ACE model, we aim to situate this conceptual contribution within the broader context of AI safety and autonomous agent research.

\subsection{The Industry Perspective}

The ACE framework emerged from observing the rapid growth of autonomous AI development in industry and open source communities. As researchers studying AI advancements, we recognized the increasing urgency to create autonomous systems capable of independently achieving goals. Tech giants compete to launch household robots and self-driving cars, while startups propose virtual assistants and self-thinking drones. Open source GitHub repositories host numerous projects on autonomous game NPCs and robotic control algorithms.

However, we observed that much progress resulted from ad-hoc experimentation rather than systematic architectural thinking. Companies combined machine learning models, hoping for autonomous performance to emerge. Hackathons produced small, incremental improvements without a comprehensive view of autonomous machines or connections to human cognition.

In response, we aimed to formalize a conceptual framework reflecting best practices for designing autonomous systems. By examining successful developers' approaches, we identified key principles around layered abstraction, integrated ethics, and human-aligned adaptation. This led to the Autonomous Cognitive Entity model - our attempt to offer blueprints for engineering autonomous AI.

Similar to how architectural and engineering principles evolved for complex modern buildings, the ACE framework provides developers with a robust architecture for autonomous cognition. As the demand for capable and beneficial autonomous AI continues, we hope these conceptual blueprints assist teams in building ethical, safe, and human-centered cognitive agents. The ACE model, derived in part from field observations, aims to address the need for structured thinking on autonomous architectures.

\subsection{Current Limitations of LLMs}

Large language models (LLMs) signify a paradigm shift in artificial intelligence, but their limitations and proper use remain debated. Although LLMs generate fluent human-like text, their understanding depth is uncertain. Some researchers claim LLMs possess human-like reasoning, common sense, and theory of mind, while others argue they exploit surface-level statistical patterns without genuine comprehension of semantics or reality grounding. This relates to broader questions of whether capabilities like reasoning and theory of mind are well-defined or measurable in machines. Proposed benchmarks for LLMs face criticism regarding validity. For example, benchmarks testing factual knowledge are limited by training datasets and don't assess knowledge integration and reasoning. Tests of narrative understanding and theory of mind are inconclusive, as LLMs can superficially imitate abilities without true comprehension. Open challenges remain in creating benchmarks that robustly characterize capacities like common sense.

Debates continue about whether external grounding or embodiment is necessary for understanding versus purely self-contained statistical learning. Some argue sensory experiences grounding is essential for semantics and generalization, while others suggest internal statistical coherence suffices for specialized applications. Resolving these theoretical disputes is challenging empirically and beyond this paper's scope. Additionally, deep philosophical puzzles persist regarding definitions of intelligence and consciousness in LLMs. These issues intersect with ethics concerning AI rights and personhood. While these philosophical questions have historical roots, LLMs present them in new forms. If an entity exhibits all objective hallmarks of intelligence and consciousness, how do we distinguish life from non-life? Many of these questions extend well beyond the scope of this paper.

\subsection{Practical Implications}

The ACE model has extensive practical implications, applicable in various domains. Integrating large language models and multimodal generative models, it can create autonomous systems capable of complex tasks, adapting to changes, and making ethically aligned decisions. In healthcare, the ACE model could develop autonomous agents assisting doctors in disease diagnosis, treatment planning, and patient health monitoring. These agents could adapt their actions based on the patient's condition, doctor's directives, and medical ethics, ensuring effective and ethical healthcare services. In cybersecurity, the ACE model could create autonomous agents monitoring network activity, detecting security threats, and responding to attacks. These agents could adapt their actions based on the threat, security team directives, and cybersecurity principles, ensuring robust and flexible security solutions.

Overall, the ACE model's extensive practical implications can revolutionize autonomous systems by integrating advanced AI technologies and insights from multiple disciplines, leading to more robust, flexible, and effective cognitive architectures.

\subsection{Comparison with other Frameworks}

A key part of assessing any new conceptual model is comparing it to existing related frameworks, analyzing the similarities, differences, and unique contributions. This section will compare the layered architecture of the proposed Autonomous Cognitive Entity (ACE) model with two alternative cognitive architectures from recent research – the Autonomous Machine Intelligence (AMI) model \cite{lecun2022path} and the Cognitive Architecture for Language Agents (CoALA) framework \cite{sumers2023cognitive}. By elucidating the key distinctions between ACE and these other approaches across each architectural layer, this analysis aims to highlight the novel aspects of ACE's design. The comparisons focus on how the frameworks differ in their structure, capabilities, and integration of components for autonomous cognition. Examining these architectural variations provides perspective into how ACE diverges from prior architectures and establishes a distinct paradigm.

\textbf{Aspirational Layer}: The Aspirational Layer is a key conceptual innovation in the ACE framework focused on establishing high-level ethical principles, values, and imperatives to guide agent behavior. In contrast, the AMI framework lacks an explicit aspirational reasoning module, with the closest analogue being the Intrinsic Cost module encoding basic drives rather than abstract ethics. The CoALA framework incorporates some intrinsic motivations and philosophical ethics to shape objectives, but its formulation is more technical than the ACE Aspirational Layer's natural language principles focused on idealized, universal morality. Overall, the distinct Aspirational Layer in ACE operates at a higher level of abstraction centered on moral reasoning rather than individual drives or technical metrics. By embedding ethics as the topmost oversight layer, ACE structurally enforces a clear separation between aspirational judgment and lower-level action, which AMI and CoALA lack. This architectural choice reflects ACE's emphasis on aligning agent behavior to human values through prioritizing ethical reasoning.

\textbf{Global Strategy Layer}: The ACE Global Strategy Layer devises high-level plans and strategies guided by principles from the Aspirational Layer, leveraging latent knowledge within language models. This bears some resemblance to AMI's World Model learning environment dynamics and CoALA's Actor proposing action sequences. However, ACE's Global Strategy Layer plays a more central role in directing behavior based on ethical oversight and long-term reasoning beyond immediate actions. It provides targeted grounding to focus the language model's imagination toward useful outcomes aligned with the agent's context and goals. In contrast, AMI and CoALA lack integrated top-down guidance, with planning modules focused narrowly on technical optimization.

\textbf{Agent Model Layer}: The ACE Agent Model Layer develops an explicit computational representation of the agent's capabilities, architecture, and limitations. This facilitates reasoning and planning based on an embedded perspective of the agent's self-knowledge. Neither AMI nor CoALA have an analogous distinct metacognitive self-modeling layer. Instead, AMI distributes related functions like skill learning and memory across modules like the Actor and World Model. CoALA's Actor selects actions based on skills learned through environmental interaction rather than internal self-modeling. The segregated Agent Model Layer in ACE provides architectural innovation in integrated metacognition and self-awareness missing from both AMI and CoALA.

\textbf{Executive Function Layer}: The ACE Executive Function Layer concretizes high-level plans into detailed actionable routines, incorporating oversight responsibilities like risk assessment and resource management. This extends beyond AMI's Actor focused narrowly on technical path planning and CoALA's Actor converting strategic objectives into incremental action steps. ACE's Executive Function Layer leverages robust inputs from upper layers for comprehensive pragmatic planning aligned with the agent's principles, objectives, and limitations. In contrast, AMI and CoALA lack strong hierarchical integration between conceptual oversight and concrete planning.

\textbf{Cognitive Control Layer}: ACE's Cognitive Control Layer implements metacognitive functions like frustration tolerance and cognitive damping for flexible decision-making, especially in uncertain or conflicting situations. Neither AMI nor CoALA incorporate explicit architectures for cognitive control. Their reactive approaches leaves them vulnerable in disruptive scenarios where core assumptions are invalidated. ACE's specialized mechanisms modeled on human cognition provide crucial resilience, enabling the agent to safely and intelligently adapt when initial plans fail. This represents a key point of differentiation from AMI and CoALA.

\textbf{Task Prosecution Layer}: The ACE Task Prosecution Layer separates basic execution from cognition, which resides in higher layers. This differs from AMI and CoALA where planning and reasoning modules are tightly coupled to embodiment. By isolating general reasoning capacities from situation-specific skills, ACE gains flexibility regarding diverse interfaces to the external world. In contrast, bundling cognition and physical skills limits AMI and CoALA's transferability across contexts relative to ACE's emphasis on platform-independent reasoning.

While ACE shares high-level similarities with AMI and CoALA, its specialized focus on ethical reasoning, metacognition, cognitive control, and transferable reasoning differentiates its layered architecture and approach to developing beneficial autonomous intelligent systems. The comparisons illuminate ACE's conceptual innovations in integrating human values, robust abstraction, and flexible adaptation within a hierarchical cognitive framework.

\subsection{Philosophical Considerations}

The ACE framework presents a novel approach to autonomous cognitive architectures. However, it is crucial to note that the full ACE model has not been implemented yet. Each architectural layer is based on existing research and industry implementations of specific capabilities. For example, the Aspirational Layer for ethical reasoning builds on AI ethics and value alignment work, while the Task Prosecution Layer for skill execution utilizes advances in robotic control and natural language processing. This paper is an initial effort to combine progress across fields into a unified architectural paradigm. The next phase involves actualizing the ACE model through incremental prototyping and comparative benchmarking against alternative approaches. We propose a methodology for rigorous, multi-faceted evaluation of future ACE implementations, but validating the framework's capabilities and benefits is ongoing future work dependent on an operational prototype system.

We present this research as an exploration of a promising design space for artificial general intelligence, rather than making definitive claims on feasibility or benefits. The ACE model introduction aims to foster future work on autonomous architectures integrating insights from neuroscience, psychology, and philosophy. 
This paper focuses on conceptual contributions rather than demonstrated benefits, situating the current work as preliminary theory development and architectural design requiring extensive practical realization and validation. Our intention is to provide the conceptual groundwork, guiding subsequent engineering efforts towards beneficial autonomous cognitive systems.

\subsubsection{The Need for Grounded Meaning}

A valid criticism of the ACE framework is its reliance on large language models (LLMs) for reasoning and decision-making, as they lack inherent understanding of truth or connections between symbols and real-world referents. LLMs reason based on statistical patterns over text corpora, without grounding in external reality or sophisticated theories of meaning. This lack of grounding can lead to false inferences, misunderstandings, and untrue statements, while enabling imaginative responses detached from objective facts. Without grounding, LLMs can hallucinate any version of reality, as long as it is statistically coherent with their training data.

This issue emphasizes the importance of context in guiding LLM reasoning. By providing relevant assumptions and goals, the latent knowledge within LLMs can be directed towards useful responses grounded in the current situation. Layers like the Global Strategy and Agent Model offer this contextual grounding. The Global Strategy Layer integrates real-time information about the agent's environment and broader context, giving the LLM key facts to reason about rather than operating in a contextual vacuum. The Agent Model Layer provides self-knowledge about the agent's capabilities and limitations, further orienting the LLM towards pragmatic responses tailored to the agent's abilities.

Together, the contextual grounding from upper layers focuses the LLM's generative capacity on productive outcomes grounded in the current circumstances and directed towards the agent's goals. Explicitly specifying the desired reasoning context is essential to beneficially leveraging the LLM's statistical imagination while avoiding unmoored hallucinations. Integrating outside knowledge to orient the LLM and rigorously verifying outputs can mitigate risks from the lack of inherent grounding in external reality.

\subsubsection{Epistemic Considerations}

The ACE framework incorporates philosophical principles to guide agent decision-making and ensure ethical alignment; however, open epistemological questions remain regarding how large language models (LLMs) represent and reason about concepts related to knowledge, truth, understanding, and meaning. Although LLMs exhibit some human-like cognitive capabilities, such as theory of mind and common sense reasoning, the underlying mechanisms are not fully understood, and the relationship between statistical patterns in textual training data and human-like conceptual knowledge remains unclear\cite{li2022emergent, chen2023beyond}.

The ongoing debate questions whether LLMs' capabilities arise from learning similar information processing strategies as humans or from fundamentally different computational mechanisms. Training on large text corpora, like humans, could potentially lead to convergent representational spaces and reasoning abilities; however, LLMs may also develop divergent techniques specialized for statistical pattern recognition that do not reflect human understanding. Assuming LLMs gain human-like "understanding" or conceptual knowledge reconstruction from statistical co-occurrence patterns is speculative, and we lack a comprehensive grasp of how LLMs generalize epistemic models beyond their training distributions. Significant gaps remain in understanding how LLMs represent abstractions related to truth, meaning, inference, and semantics. Indeed, we do not fully comprehend human generalization of understanding!

While LLMs demonstrate potential in replicating aspects of human intelligence, we must exercise caution against prematurely concluding that they fully capture complex philosophical notions underpinning human cognition. Further interdisciplinary research is required to thoroughly assess the epistemic capacities and limitations of large language models in frameworks like ACE.

\subsubsection{Known Gaps and Assumptions}

The ACE framework integrates insights from diverse fields like neuroscience, psychology, philosophy, and computer science, but significant gaps in understanding within these disciplines necessitate making assumptions. Human cognition provides limited insights into consciousness, theory of mind, and other complex mental faculties. Although the ACE framework incorporates current theories, much remains unknown about the human brain's mechanisms underlying mind and subjective experience. Assumptions must be made regarding similarities between ACE's cognitive layers and corresponding brain systems, but precise neuro-cognitive mappings are unclear.

In computer science, the representational capacities and limitations of artificial neural networks and large language models are not fully characterized. While they demonstrate certain human-level abilities, their internal workings are not well understood. It is uncertain how mathematical embeddings might translate to conceptual knowledge or if different computational mechanisms are involved. The ACE framework assumes sufficient commonality to human cognition for insight transfer.

From a philosophical perspective, open questions persist regarding ontology, meaning, truth, consciousness, and other domains. The ACE framework strives for conceptual balance but adopts a functionalist approach focused on developing beneficial autonomous systems. For example, both deontological and teleological ethics are integrated based on their complementary utility rather than assertions about metaphysical reality, acknowledging the limitations in digitally instantiating abstract philosophical notions.

Realizing the ACE vision requires making assumptions regarding gaps in current understanding at the frontiers of neuroscience, artificial intelligence, and philosophy. As research progresses, these gaps will incrementally narrow, allowing for ACE framework refinement to better match human-level cognitive capabilities. The current model represents the best synthesis given the available knowledge across these complex and interdisciplinary topics.

\subsubsection{Model Dependent Ontology}

It is worth noting that some philosophical perspectives argue external grounding may not be strictly necessary for language and reasoning to function effectively in artificial systems, even if it departs from human cognition. For instance, the epistemic framework of Model Dependent Ontology (MDO) \cite{Delaflor2023intro}, could offer an alternative foundation for a more advanced ACE architecture in the future. This framework posits that large language models demonstrate we do not necessarily require external "ground truth" references for language to cohere within a closed conceptual system. Rather than relying on conventional realist assumptions behind human cognition, MDO illustrates an approach focused on internal consistency and usefulness over correspondence to an imposed external world.

Specifically, Model-Dependent Ontology affects knowledge representation in artificial agents by emphasizing flexibility in conceptual modeling unbound by assumptions of a single objective reality. It allows coexistence of multiple valid yet incompatible models of phenomena based on differing internal assumptions. Additionally, MDO decouples models from physical constraints, enabling exploration of purely conceptual spaces detached from sensorimotor limitations. This framework judges models primarily based on their internal coherence and usability rather than accuracy to external stimuli. The emphasis is on developing maximally useful representations for a given context rather than objectively true representations. Another form of grounding can be found in contextual references. For instance, using several layers on the ACE helps to keep hallucinations under control by enhancing the context to more than one layer.

By relaxing realist assumptions, MDO opens possibilities for artificial systems to generate and leverage speculative conceptual models that diverge from human consensus reality. Within this paradigm, agents can develop their own optimal conceptual symbols and ontologies without needing to ground them in a predefined external world. In essence, MDO takes a pragmatic engineering approach focused on what forms of reasoning work effectively rather than adhering to philosophical ideals of truth and grounded meaning derived from human cognition.

This alternative perspective indicates external grounding, while critical for human-level understanding, may not be an absolute requirement for artificial systems to operate effectively in specialized niches. The flexibility and internal coherence enabled by model-dependent reasoning suggest further exploration of non-grounded approaches could yield useful technological systems capable of reasoning in ways departing from biological cognition. As such, the merits and limitations of both grounded and non-grounded paradigms remain open research questions worthy of continued investigation within the ACE framework and artificial intelligence more broadly.
 
\subsection{The Path Forward}

The growing presence of autonomous AI systems in industry highlights the need for increased academic involvement to incorporate ethical and philosophical perspectives into their development. By contributing frameworks like ACE, researchers can help guide the development of autonomous AI towards a beneficial direction. However, fully actualizing the ACE model as a mature architectural paradigm necessitates extensive future research.

One crucial direction is developing detailed reference architectures, specifications, and standards based on the high-level ACE framework. Organizations like IEEE could serve as a model for rigorously defining key aspects of the ACE layers, interactions, and interfaces. Concrete canonical instantiations would expedite the translation of the conceptual ACE model into functional implementations. Ongoing research and debate are essential for addressing philosophy, ethics, values, and aligning autonomous systems with human needs. Initiatives like AI4People foster discussions on utilizing AI to promote human dignity and rights. Collaborative forums can help guide development towards human-flourishing outcomes by further defining beneficial AI.

Empirical research is vital for evaluating implementations, capabilities, and limitations. Real-world testing through benchmark tasks and experimental deployments will reveal strengths and areas for improvement. Developing rigorous benchmarks that avoid pitfalls like anthropic biases observed in some previous language model tests is a priority. Human-centered design insights can also inform the user experience of autonomous systems. Evidence-based research can refine the ACE framework over successive iterations, systematically progressing towards artificial general intelligence centered on human needs.

The primary path forward involves implementing and evaluating the ACE framework in applied autonomous software, revealing its strengths and weaknesses through real-world testing and iterative refinements. Benchmarking and comparing alternative cognitive architectures will highlight the merits and limitations of the ACE approach. Continuously improving and evaluating core software components, particularly large language models, will enhance ACE-based systems' capabilities. However, the framework is model agnostic, focusing on architectural principles rather than specific machine learning techniques, encompassing a broader design space for autonomous cognition and software engineering.

Realizing ACE's potential as a beneficial autonomous software architecture depends on extensive practical implementation, benchmarking, and refinement driven by real-world engineering needs. This applied systems-focused process will reveal more about layered cognitive architectures' prospects and limitations for autonomous agents compared to alternative approaches, ultimately advancing the field.

\section{Conclusion}

This paper introduced the Autonomous Cognitive Entity (ACE) framework, a novel model for artificial general intelligence based on a layered cognitive architecture. The ACE framework integrates insights from neuroscience, philosophy, psychology, and computer science to enable autonomous systems to make flexible, adaptive decisions aligned with ethical principles. The core innovation of the ACE model is its hierarchical structure incorporating six layers, each with distinct functions spanning from moral reasoning to task execution. The upper Aspirational Layer and Global Strategy Layer embed philosophical ideals and high-level planning, guiding the system's overarching direction. The mid-level Agent Model, Executive Function, and Cognitive Control Layers handle self-monitoring, dynamic task management, and decision-making. Finally, the bottom Task Prosecution Layer interacts with the environment to carry out actions.

The layered abstraction provides clear delineation between different facets of cognition while enabling bidirectional information flow. The Aspirational Layer monitors system activity through read access to all layers, allowing top-down intervention. Feedback from lower layers propagates upwards, guiding adaptation of strategic plans and ethical frameworks based on experience. Together, the six layers enable autonomous goal setting, planning, adaptation, task switching, and ethical reasoning within a single architecture. By combining abstract reasoning and concrete execution, the ACE framework provides a path toward artificial general intelligence that aligns decisions and actions with human values.

The introduced conceptual model proposes a foundation for future development of ACE systems. Potential research directions include formal verification of system properties, detailed computational implementations, and evaluation across diverse real-world deployments. As a transdisciplinary synthesis, the ACE framework underscores the importance of unifying perspectives from ethics, cognitive science, and computer engineering to create capable and beneficial autonomous agents.

\bibliographystyle{ACM-Reference-Format}
\bibliography{sample-base}

\end{document}